%% file: arxiv.tex
\RequirePackage{fix-cm}
\RequirePackage{amsmath}
\documentclass{svjour3}                     

\smartqed  

\usepackage{latexsym}
\usepackage{placeins}
\usepackage{graphicx}
\usepackage{epsfig}
\usepackage{color}
\usepackage[utf8]{inputenc}
\usepackage{amssymb}
\usepackage{amsmath}
\usepackage{stmaryrd}

\newcommand{\be}{\begin{equation}}
\newcommand{\ee}{\end{equation}}
\newcommand{\ba}{\begin{eqnarray}}
\newcommand{\ea}{\end{eqnarray}}

\journalname{Journal of Engineering Mathematics}

\begin{document}
\title{Ferrofluids and magnetically guided superparamagnetic particles in flows: A review of
simulations and modeling\thanks{This work is partially supported by NSF-DMS-1311707 and  NSF-CBET-1604351.}  }
\author{Shahriar Afkhami \and Yuriko Renardy}
\institute{Shahriar Afkhami \at Department of Mathematical Sciences, New Jersey Institute of Technology, Newark, NJ 07102 USA\\
\email{shahriar.afkhami@njit.edu}
\and
Yuriko Renardy \at
Department of Mathematics, Virginia Polytechnic and State University, Blacksburg, VA  24061 USA\\
\email{renardy@vt.edu}
}
\date{Received: date / Accepted: date}
\maketitle
\begin{abstract}
\input{abstract2}

\end{abstract}

%

\section{Introduction}
\label{intro}
\input{introduction3}
\section{Mathematical modeling of magnetic drug targeting}
\label{mdt}

\input{superparamagnet2}

\FloatBarrier
\section{Mathematical modeling of ferrofluid drops}
\label{drop}
\input{NS2}

\FloatBarrier
\input{numerics2}

\FloatBarrier

\section{Conclusion}
\input{conclusion1}
\FloatBarrier



\end{document}

%% file: abstract2.tex
Ferrofluids are typically suspensions of magnetite nanoparticles, and behave as a homogeneous continuum.  The production of nanoparticles with a  narrow size distribution and the achievement of colloidal stability are important technological issues.  The ability of the ferrofluid to respond to an external magnetic field in a controllable manner has made it emerge as a smart material in a variety of applications, such as
seals, lubricants, electronics cooling, shock absorbers and adaptive optics. Magnetic nanoparticle suspensions have also gained 
attraction recently in a  range of  biomedical applications, such as cell separation, hyperthermia, MRI, drug targeting and cancer diagnosis.
 In this review, we provide an introduction  to mathematical modeling of three problems:  motion of  superparamagnetic nanoparticles in magnetic drug targeting, the motion of a ferrofluid drop consisting of chemically bound nanoparticles without a carrier fluid, and the breakage of a thin film of a ferrofluid.

%% file: introduction3.tex
  This review presents three aspects of the mathematical analysis of the motion of superparamagnetic particles and drops in a surrounding fluid, under the 
influence of an external 
magnetic field: drug targeting, ferrofluid drop deformation, and instabilities and dewetting of a thin ferrofluid film. 
 These problems are important for a number of biomedical applications.
  The motion of a magnetic particle is determined by a force balance of the magnetic force, drag, and stochastic forces. The motion of a ferrofluid drop is 
affected by its deformation. 
  Simulations  using the Volume-of-Fluidk method are presented. Once a drop reaches the target and coats it, a thin film approximation yields insight into 
instabilities and breakup.    We refer the
reader to  recent publications, which in turn contain more  comprehensive references and historical reviews.

An advantage of magnetic drug targeting is the delivery of drugs 
to  specific tissues via  the blood vessels as pathways under  externally  
applied magnetic fields. Progress towards  robust biomedical applications  
 relies on the synthesis of appropriately coated nanoparticles, clusters 
 and ferrofluids \cite{Liu2007,ATRRWPR,Bala2014,Dung2017}.  Pros and 
cons of 
directing the magnetic nanoparticles in blood vessels and tissues are 
described in 
\cite{Voltairas02,Neuberger2005,Buzea2007,Berry2009,Mishra2010,Nacev2011}. Recent reviews 
highlight the use of magnetic 
nanoparticles in many biomedical applications; for instance,  drug targeting, magnetic fluid hyperthermia, 
   design of devices to measure targeting efficiency, and tissue engineering
\cite{Ito2005,Puri2014,Maguire2014,aljamal2016,Mohammed2017,Radon2017}. 

\begin{figure}
\includegraphics[width=1\textwidth]{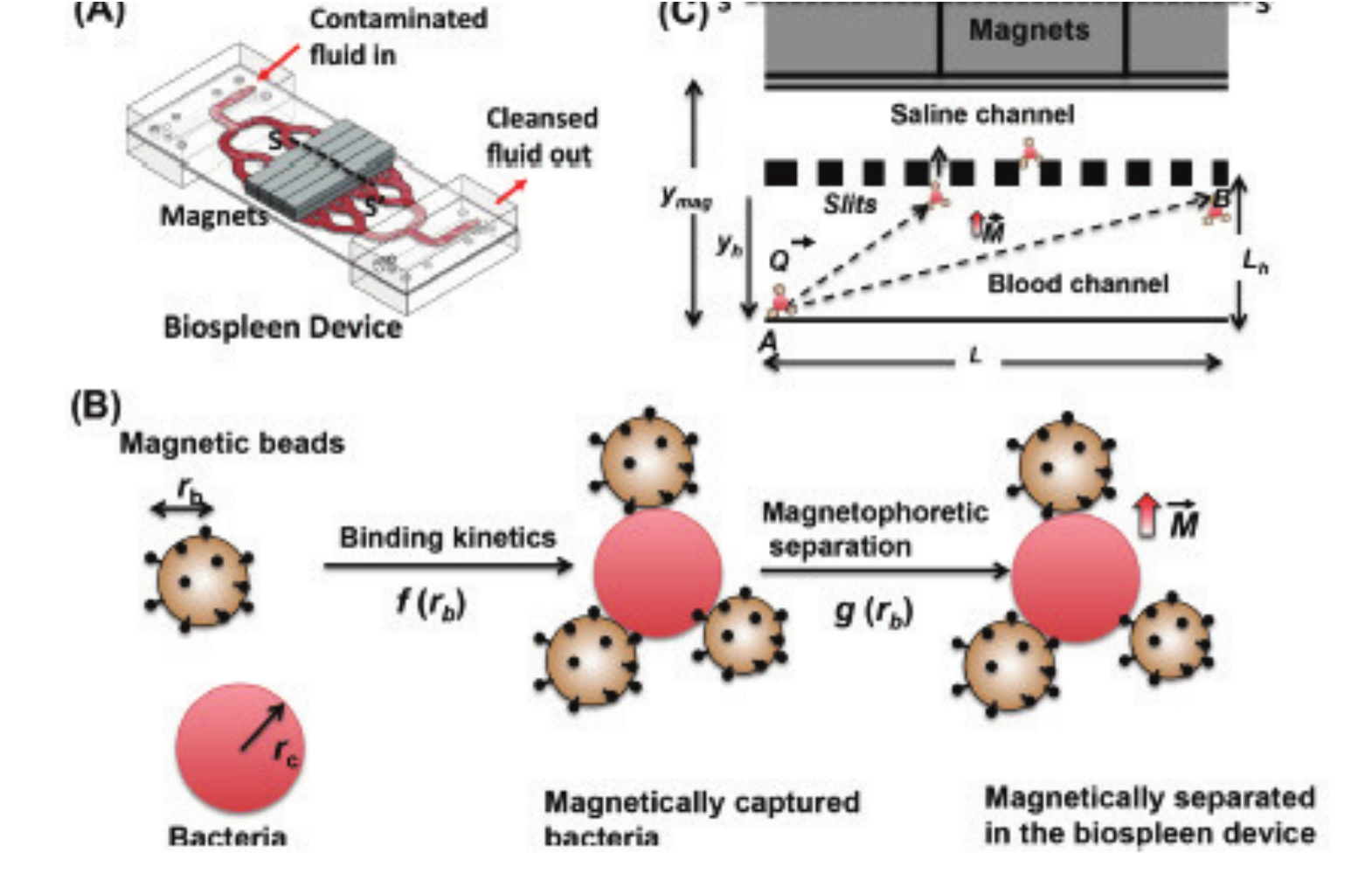}
\caption{With permission from \cite{small}. Summary of caption:  `(A) A schematic design of the biospleen that magnetically cleanses fluids contaminated with 
bacteria using magnetic nanoparticles coated with FcMBL. (B) The magnetic separation process dissociated into two sequential phases; magnetic beads binding to cells 
and magnetic separation of cell - magnetic bead complexes.  (C) A cross-sectional view of the biospleen channel through which magnetic bead-bound bacteria flow while 
they are deflected upward  by external magnets on the top. If bacteria that passed position A reach the other side of the channel before they pass position B, we can 
assume that they are completely removed.'
}
\label{kangfig}\end{figure}

In \cite{Suh2011}, the mathematical modeling of the transport of 
paramagnetic particles 
in viscous flow is  described in terms of dipole points in an external 
magnetic field, with particle interactions for  larger particles and 
higher concentrations\cite{Banerjee2012}. Numerical 
approaches using Brownian dynamics for particle interactions in 
aggregation and disaggregation are being developed \cite{vanReenen2014}.
Technological applications  include the sorting of not only cells but proteins and other biological components by biochemically funcionalized paramagnetic beads that 
are manufactured to be attracted to specific targets. The mixture flows in a fluid, and as they move through a channel, magnetic fields deflect the desired cells to 
migrate toward predetermined exits.
In \cite{Tsai2011}, paramagnetic beads floating in a liquid are subjected to a magnetic field normal to the flow. The deflection by the magnetic field steers the 
beads towards a target whose location depends on the field and on the properties of the bead.  An application to removing pathogens from the blood stream is shown in 
\cite{small}, reproduced in figure \ref{kangfig}.  Here the magnetic particles bind to pathogens such as bacteria. In addition to the motion of the 
particles, a model for the kinetics of this binding mechanism is formulated.

\begin{figure}
\begin{center}
 \includegraphics[width=0.45\textwidth]{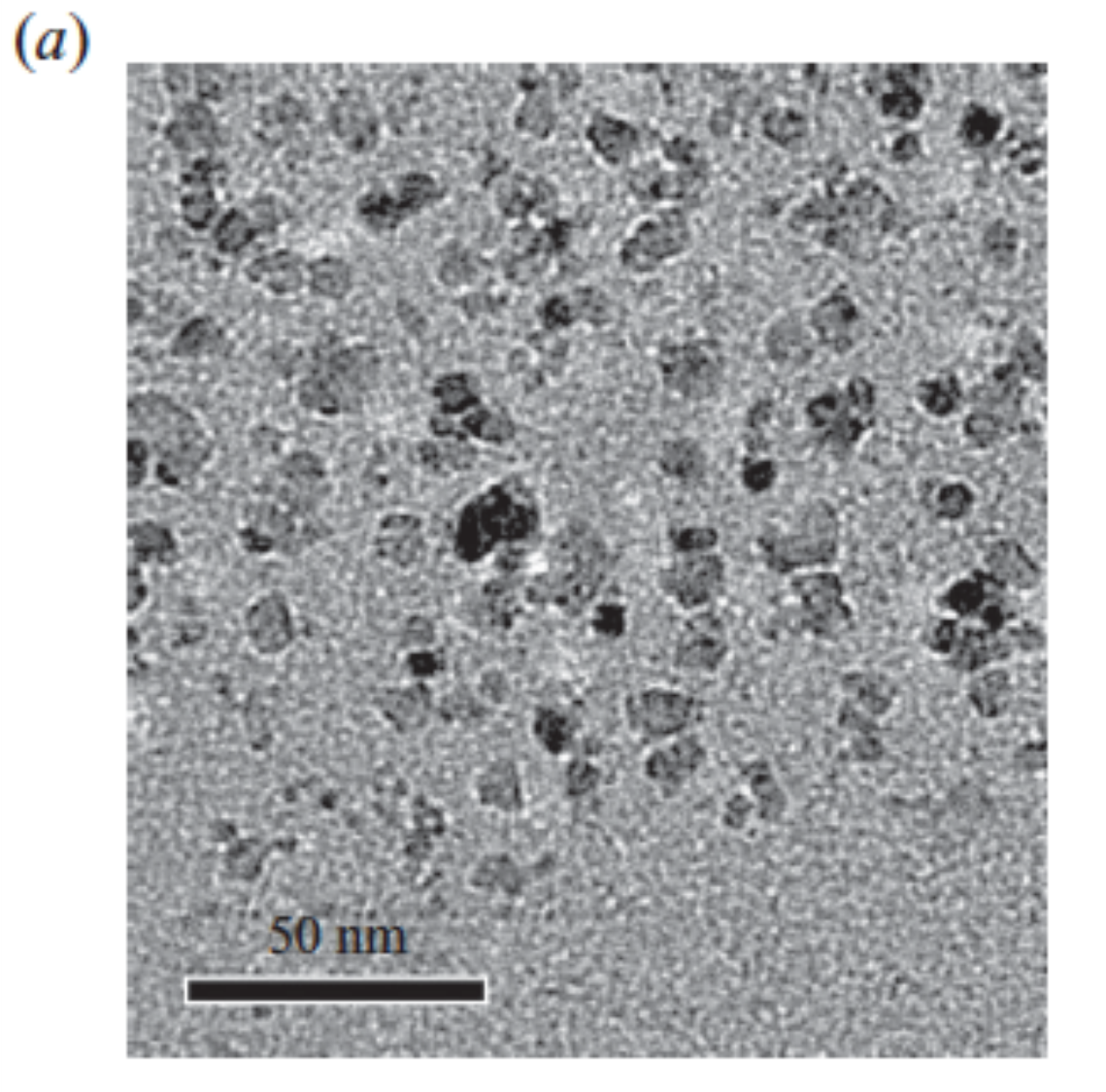}
  \includegraphics[width=0.5\textwidth]{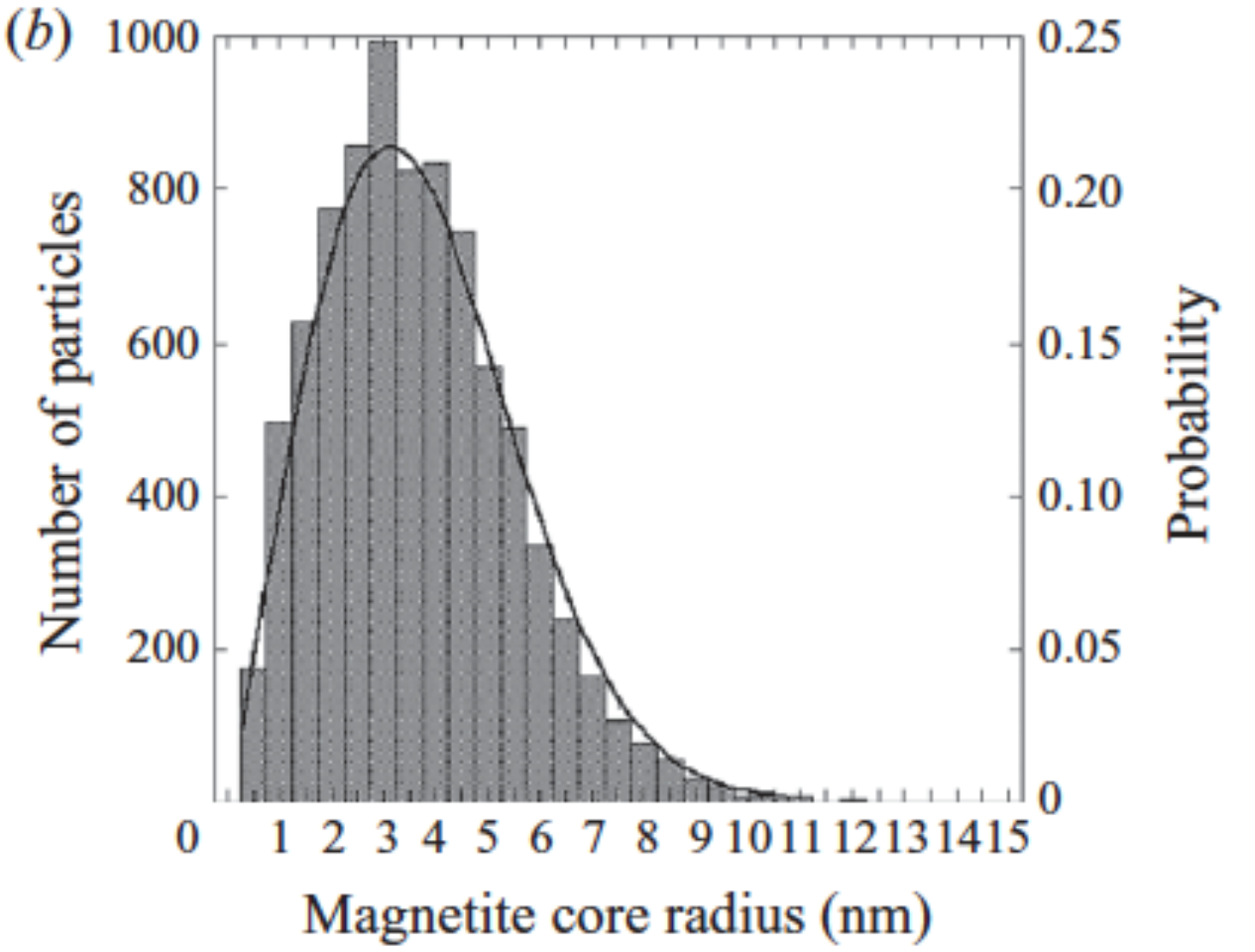}\\
(c)\includegraphics[width=0.45\textwidth,angle=90]{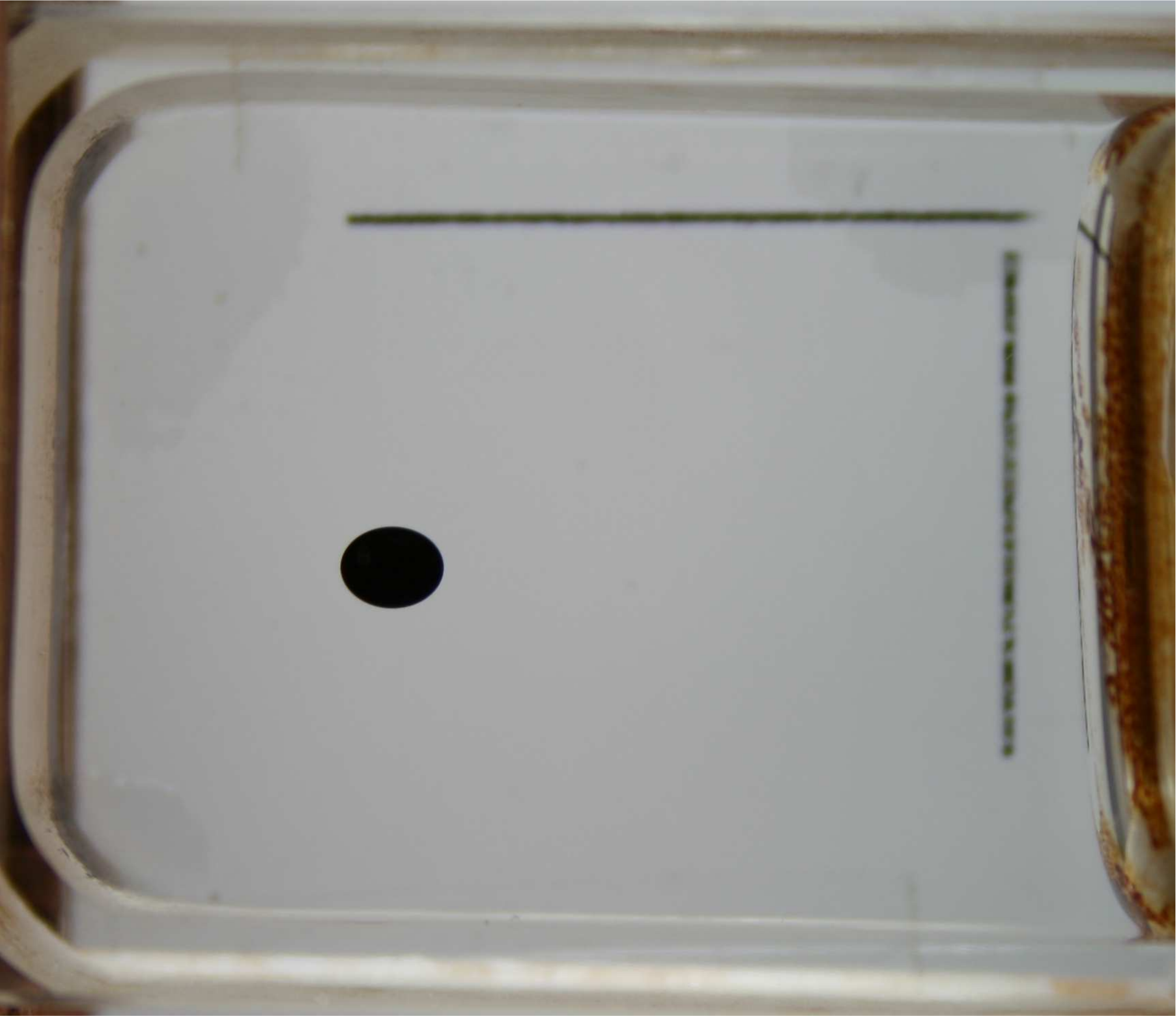}
(d)\includegraphics[width=0.45\textwidth,angle=90]{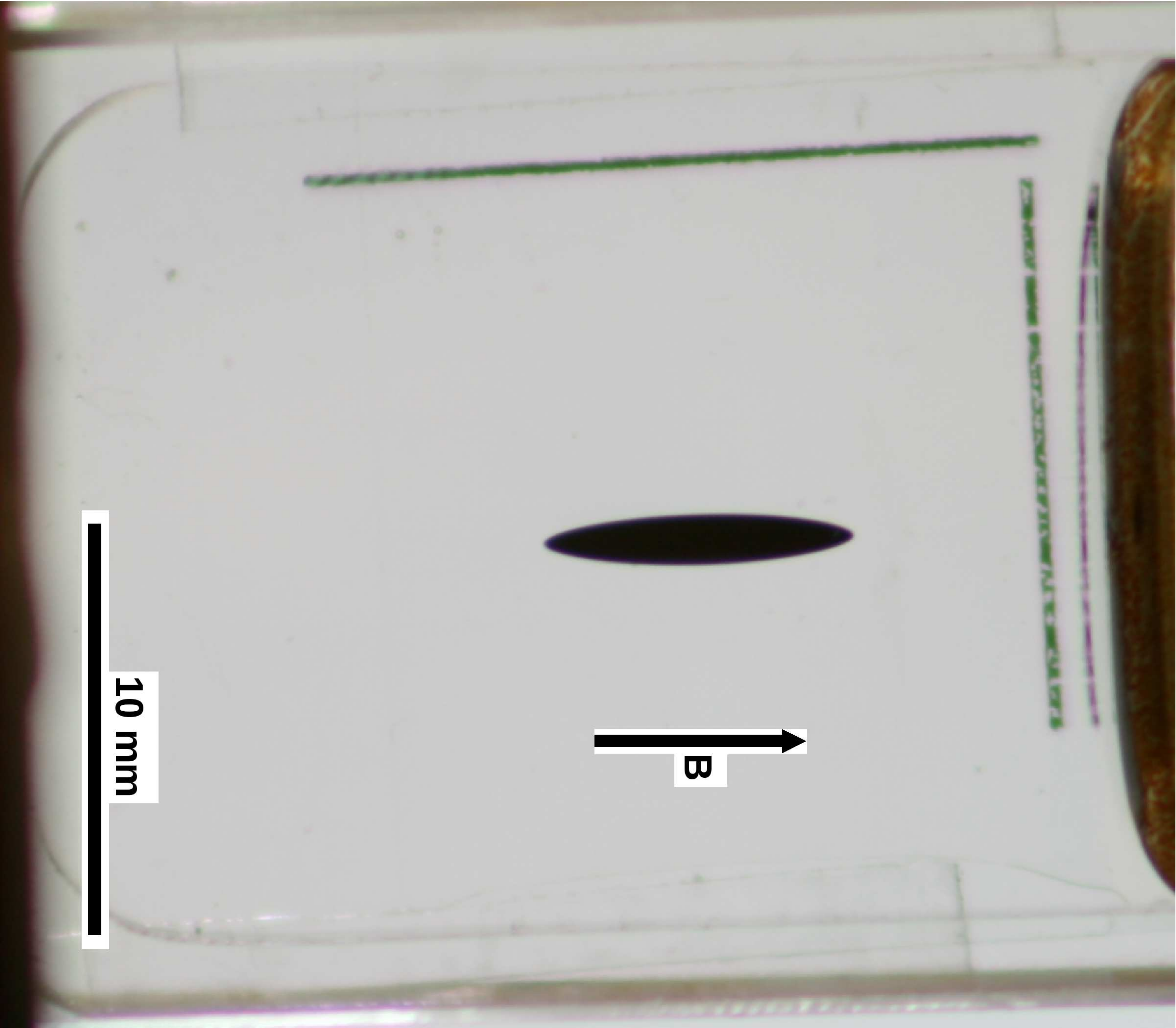}
\end{center}
\caption{(a) Representative TEM image of PDMS--magnetite particles. 
(b) Histogram of the radii of the magnetite cores. From \cite{Mefford} and \cite{Mefford2008a}.
Experimental photographs of the deformation of a PDMS ferrofluid drop
in a viscous medium under applied uniform magnetic fields of $6.38$ kA~m$^{-1}$ (c) 
and $638.21$ kA~m$^{-1}$ (d). From \cite{ARRRP} with permission.} 
\label{fig:fig1}
\end{figure}

\begin{figure}
\begin{center}
\includegraphics[scale=3]{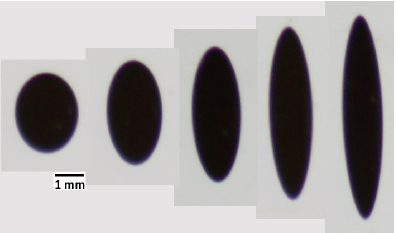}
\caption{Photographic images from \cite{ARRRP} of the drop equilibrium shape taken at
magnetic field strengths $6.032, 12.167, 23.947, 59.762, 162.33$ kA~m$^{-1}$ (from left to right).} 
\label{fig:figure12}
\end{center}
\end{figure}

The design of the superparamagnetic nanoparticles involves uniformity in the small size. Of particular interest is the formation of a ferrofluid from chemically bonding many of them. Unlike the colloidal suspension of magnetite particles, this ferrofluid avoids the  migration of the particles in the suspending fluid at high fields. This advantage is seen in the works of \cite{Mefford,Mefford2008a,Mefford2008b} for the potential treatment of retinal detachment by magnetically guiding a ferrofluid drop to seal the site. In particular, figure \ref{fig:fig1}(a) shows a sample transmission electron microscopy (TEM) 
image of the magnetite nanoparticles at $300000$ times magnification, and (b) shows the narrow size distribution. These particles are relevant to section~\ref{mdt}.  
The magnetic nanoparticles are coated with  biocompatible polydimethylsiloxane (PDMS) oligomers, and bonded to manufacture a ferrofluid. Figures \ref{fig:fig1}(c-d) show  experimentally observed states for the ferrofluid drop suspended in glycerol in uniform magnetic fields.  
The modeling in section \ref{sec:vof} concerns the numerical simulation of stable equilibrium configurations under applied uniform magnetic fields. A series of experiments conducted under low and high magnetic field strengths 
are presented in \cite{ATRRWPR} to understand the behavior of PDMS
ferrofluid drops in uniform magnetic fields. 
Figure~\ref{fig:figure12} shows the experiments in \cite{ATRRWPR} of the drop at equilibrium for
the magnetic field strengths  varying from $6.032$ to $162.33$ kA~m$^{-1}$.
These results are revisited in section~\ref{sec:vof}. For an updated review of literature
on ferrofluid drop deformation in uniform magnetic fields, see \cite{Rowghanian}.

The numerical investigation of the motion of a ferrofluid drop may be complicated by the highly distorted  material interface. An advantage of 
the boundary integral method for  three-dimensional (3D) drop deformation in Stokes flow is that the three-dimensionality is turned into a  two-dimensional surface 
computation; this together with a high order surface discretization has been able to track viscous drops with high curvature, up to the first pinch-off
\cite{CristiniBLSG,Janssen2008}. However,  this  method  does not extend easily to
non-Newtonian models such as a power-law model that is suggested for blood flow. Novel nonsingular boundary integral methods are improving the accuracy and numerical 
stability of the 3D drop deformation in viscous flow at low inertia, including the influence of a magnetic field for constant  magnetic susceptibility  
\cite{Bazhlekov06}. The effect of rotating magnetic field is investigated with the boundary integral formulation in \cite{Erdmanis2017}. 
The study of  rotating magnetic fields on ferrofluid drops is of importance in many fields including  astrophysics, and we refer the reader to recent reviews 
\cite{Lebedev2003,Fengchen2016}. A limitation of the boundary integral formulation is that it is no longer applicable if the magnetic behavior is nonlinear.

Numerical algorithms with a diffuse-interface method have been developed  
 for a droplet or layer in a uniform and rotating magnetic fields  \cite{Fengchen2016}. Intricate equilibrium shapes arise in the Rosensweig instability which 
originally refers to a ferrofluid layer under magnetic force, surface tension force, gravity and hydrodynamics \cite{Cowley,Rosensweig,Lange2007,Kadau2016}. In 
\cite{Lavrova2012}, the Rosensweig instability is investigated numerically for the diffusion of magnetic particles. The magnetic particle concentration and the free 
surface shape are part of the solution. A finite element method is used for Maxwell's equations and a finite difference method is used for a parametric 
representation of the free surface. 
 A finite element formulation is developed and used in \cite{Kang2013} to simulate the interaction of two or more particles in uniform and rotating magnetic fields. 
The interaction is  significant in  dense suspensions or larger sizes.   In section \ref{sec:vof}, we focus on  a 
Volume-of-Fluid formulation of \cite{ARRRP} which is 
capable of utilizing a non-constant susceptibility,  time-dependent evolution,  and  interface breakup and reconnection.

Finally, thin films driven by a magnetic field have been studied recently \cite{Seric2014,Conroy2015}. 
These models are derived for the flow of a thin ferrofluid film on a substrate in the same 
spirit as the previous work by Craster and  Matar \cite{Craster2005} for the case of a leaky 
dielectric model. The derivation for a nonconducting ferrofluid simplifies since there is no interfacial
charge involved. The work of Seric et al.~\cite{Seric2014} includes 
the van der Waals force to impose the contact angle using the disjoining/conjoining pressure model
and studies the film break up and the formation of satellite (secondary) droplets on 
the substrate under an external magnetic field. In section \ref{sec:lub}, we focus on  
a thin film approximation of ferrofluid on a substrate subjected
to an applied uniform magnetic field and present the results of 
the magnetic field induced dewetting of thin ferrofluid films and the consequent 
appearance of satellite droplets. 
\FloatBarrier

%% file: superparamagnet2.tex
\begin{figure}[t]
(a){\includegraphics[width=0.8\textwidth]{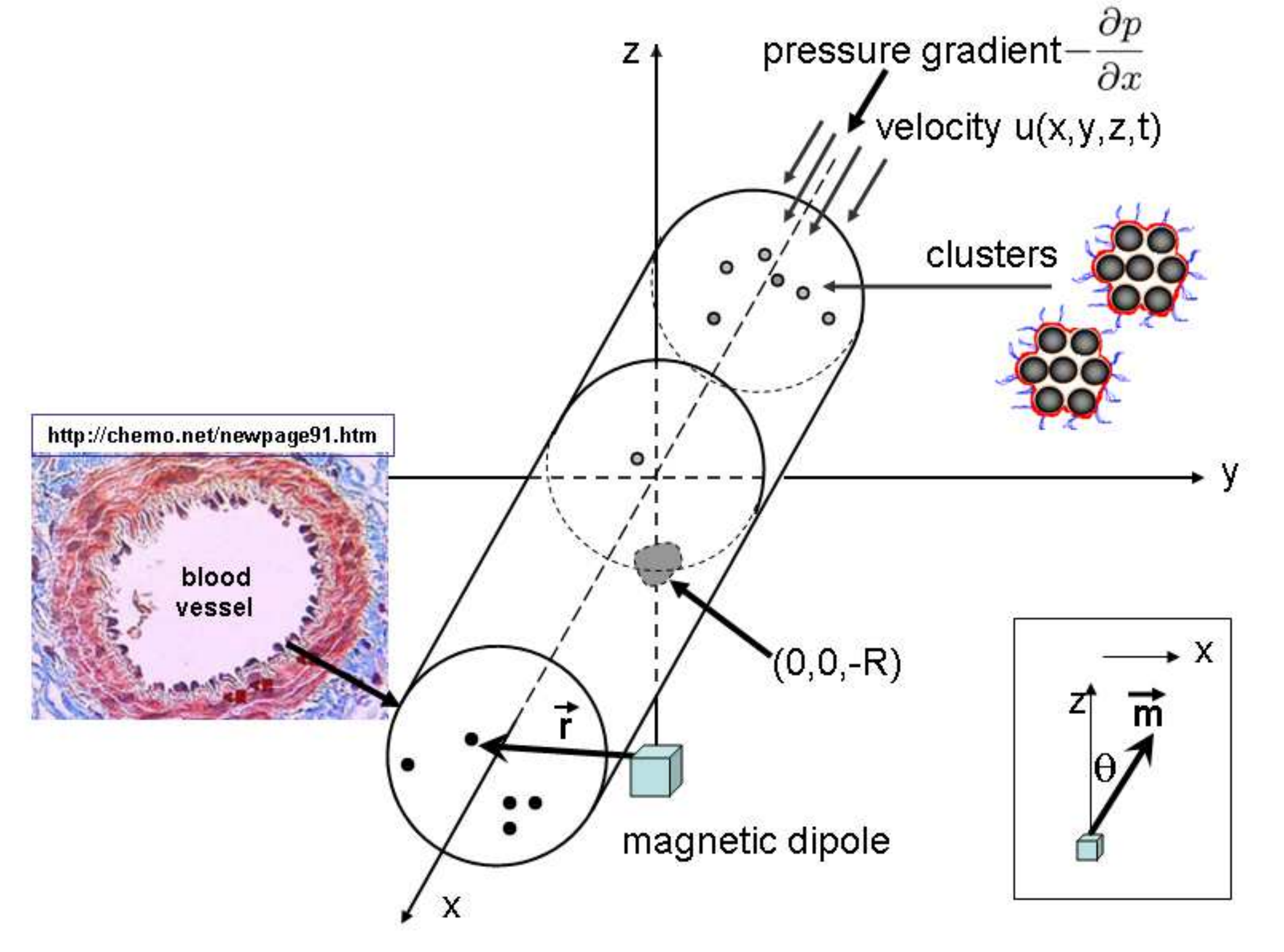}}\hfil\\
(b)\includegraphics[width=0.8\textwidth, trim = 1cm 4cm 1cm 7cm, clip]{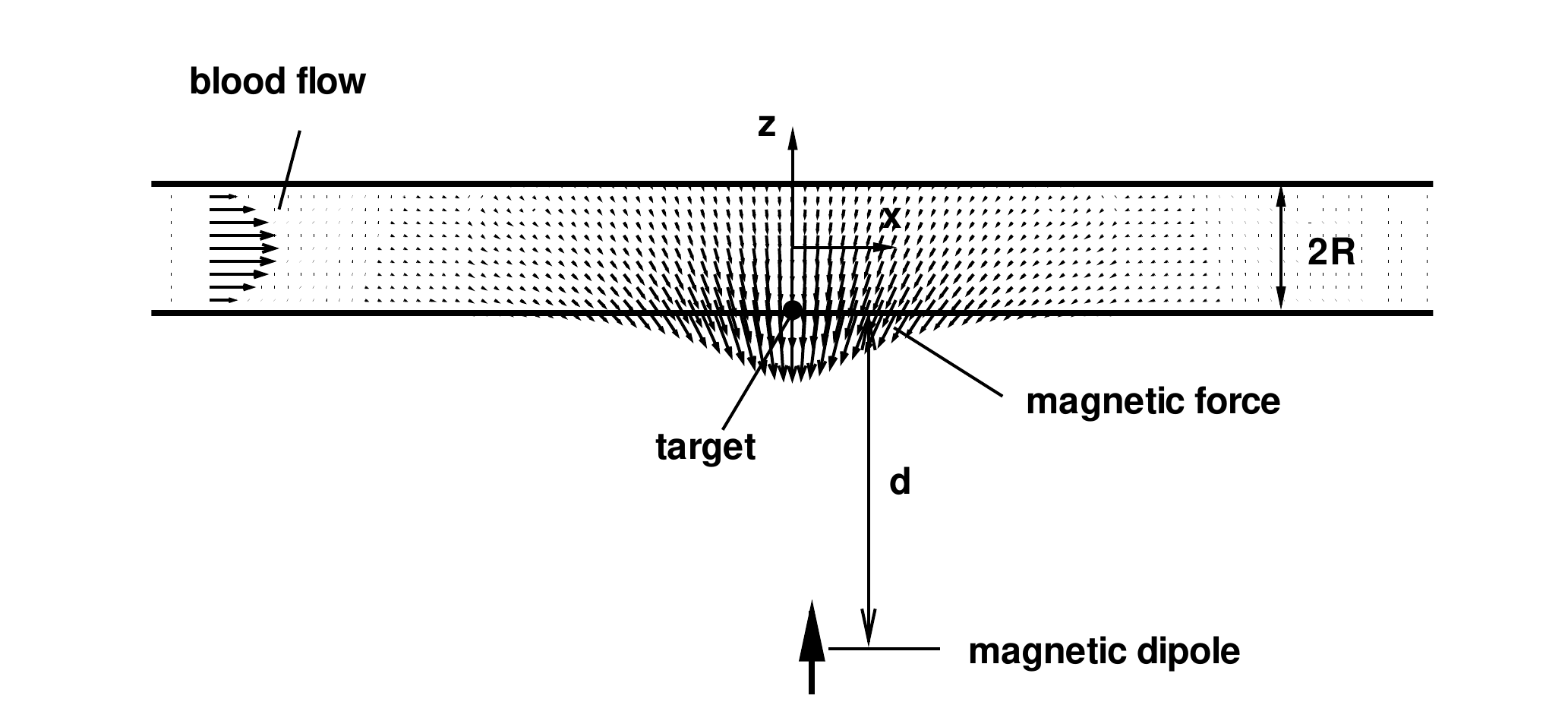}\hfil\\
\caption{(a)  (Figure 1 of \cite{YLAR2012})  The target is  at $(0,0,-R)$, 
in a cylindrical blood vessel of radius $R$.    A  magnetic point dipole  is defined by the dipole moment 
vector $\vec m$, directed from the magnet placed  below the target, and at an angle 
$\theta$ to the positive x-axis;  i.e.,  $\theta=0$ is upward.   The cross-sectional slide  
is from \cite{arterial}. 
(b) (Figure 6 of \cite{YLAR2012}) Schematic of the vertical slice through the cylindrical vessel at $y=0$. The magnetic force field
is shown when the dipole is at a
distance $d$ below the vessel wall. }
\label{schematic}
\end{figure}

In this section, we summarize the set-up and results of \cite{YLAR2012}.
One of the simplest  model problems for predicting the 
accuracy of drug targeting concerns superparamagnetic nanoparticles of at 
most several nanometers diameter, and flowing through a  venule of order 10$^{-4}$~m radius; see 
 figures \ref{schematic} (a - b). The  velocity is  $(u(y,z,t),0,0)$ where $u$ denotes the axial component.  The no-slip conditions at the
walls $y^2+z^2=R^2$ and the Stokes equation are applied.   For such a small radius, the effect of cardiac pumping on pressure 
fluctuations is negligible; therefore, $u=u_{max}(1-\frac{r^2}{R^2})$, where the flow rate $\frac{\pi}{2}u_{max}R^2$ is typically $10^{-10}$m$^3$s$^{-1}$ \cite{House86}. 
This 
gives $u_{max}\approx 0.006$m s$^{-1}$.  This 
flow generates a Stokes drag force on a solid sphere: ${\bf F}_v=-(6\pi\eta a)(\frac{d{\bf x}}{dt}-(u,0,0))$ where $\eta$ is the viscosity, $a$ the radius, ${\bf x}$ is the location of  the particle, and  $D$ is the friction coefficient. For a larger blood vessel, the oscillatory component of the flow can be estimated from the systolic and diastolic blood pressure data. 

 The external magnetic field ${\bf H}_e$  without a particle is known from experimental measurements and is used to calculate the field in the presence of the particle.  Under the  assumptions of  
   constant susceptibility $\chi$ (magnetic field not too high),  the  permeability is $\mu=\mu_0(1+\chi)$, where $\mu_0$ denotes the permeability of vacuum. Maxwell's 
equations (curl ${\bf H}_e={\bf 0}$, div ${\bf H}_e=0$) give  ${\bf H}_e=\nabla\phi$, where $\phi$ is a harmonic function. 
The external magnet is  modeled by a magnetic dipole. We find the actual field ${\bf H}$ with the particle by using the known formula $\phi=\frac{-1}{4\pi}
\frac{ {\bf m}\cdot  {\bf r}}{r^3}$, where the vector ${\bf r}$ points from the dipole to the 
particle.  Provided that the external magnet is small,  it is approximately a point dipole.
This magnetic field exerts a force on the  particle:  ${\bf F}_m=\int {  \mu_0 ({\bf M}\cdot\nabla){\bf H}_e  dV}$ where ${\bf M}=\chi{\bf H}$ denotes the magnetization corresponding to 
the magnetic field ${\bf H}$ with a particle of volume $V=\frac{4}{3}\pi a^3$, namely the Clausius-Mossotti formula 
\cite{Rosensweig,Cohen2007} \ 
 ${\bf M}=\frac{3\chi}{3+\chi}{\bf H}_e$. This is used to calculate ${\bf F}_m$. The governing equation  must satisfy ${\bf F}_m+{\bf F}_v={\bf 0}$. This is a system of nonlinear ordinary differential equations.

  The superparamagnetic particles and clusters addressed in \cite{YLAR2012} are small enough that the N\'eel relaxation time and the  magnetization are  instantaneous compared with the motion to be solved\cite{Bala2014}. However, we need  to keep in mind the that Brownian motion  should be included  if thermal energy $kT$ becomes  comparable to magnetic energy $\mu_0 MHV$, where $k$ denotes Boltzmann's constant, $T$ is in degrees Kelvin, $H$ is the field and $V$ is the volume of the particle.  This estimate gives the diameter to be smaller than  $(6kT/\pi\mu_0MH)^{1/3}$  ((2.2) of  \cite{Rosensweig}). In the examples  below, this is diameter of order $10^{-8}$m. 
 We incorporate Langevin's model for  Brownian motion 
through a stochastic forcing vector.  This leads to the system
\begin{equation}
\rho V\frac{d^2 {\bf x}}{dt^2} = -D (\frac{d{\bf  x}}{dt} - {\bf u_b}) + \frac{3 \chi V \mu_0  }
{ 3 + \chi} \nabla \left(\frac{1}{2}|\nabla\phi|^2\right) + \frac{\sqrt{2DkT}}{\sqrt{{\bf d}t}}{\bf N}(0,1), \label{brownian}
\end{equation}
where $\rho$ is the density of the particle/cluster, and ${\bf u_b}$ is
the base flow.  The left hand side term denotes acceleration which is estimated to be relatively small. $N_1(0,1)$, $N_2(0,1)$ and $N_3(0,1)$ denote independently generated, normally distributed,
random variables with zero mean and unit variance . The random variables $N_i(0,1)$, $i=1,2,3,$ are  constants over
very short time intervals $dt$, and change randomly with a Gaussian distribution. Finally, the stochastic ordinary differential equation for the state vector
$X (t)= {\vec x}^T =(x(t),y(t),z(t))^T$ is,
\begin{eqnarray}
dX && =f(X) dt + g dW, \\ 
f(X) && = \left[{\bf u_b}({\bf  x},t)) +  \frac{3 \chi V \mu_0  }{ (3 + \chi) D} \nabla \left( \frac{1}{2} |\nabla\phi(\vec x)|^2 
\right) \right]^T, \\
g && =  \sqrt{\frac{2kT}{D}} \mathbf{I};\label{system2}
\end{eqnarray}
 $\mathbf{I}$ is the $3\times 3$ identity matrix,
$f(X(t))$ is the $3\times 1$ drift-rate vector, $g$ is a  $3\times 3$ instantaneous diffusion-rate
matrix, and $dW(t)$ denotes the vector  $\sqrt{{\bf d}t} {\bf  N}(0,1)$.  

\subsection{Numerical algorithm}
\begin{figure}[t]
\includegraphics[width=0.45\textwidth,angle=-90,origin=c]{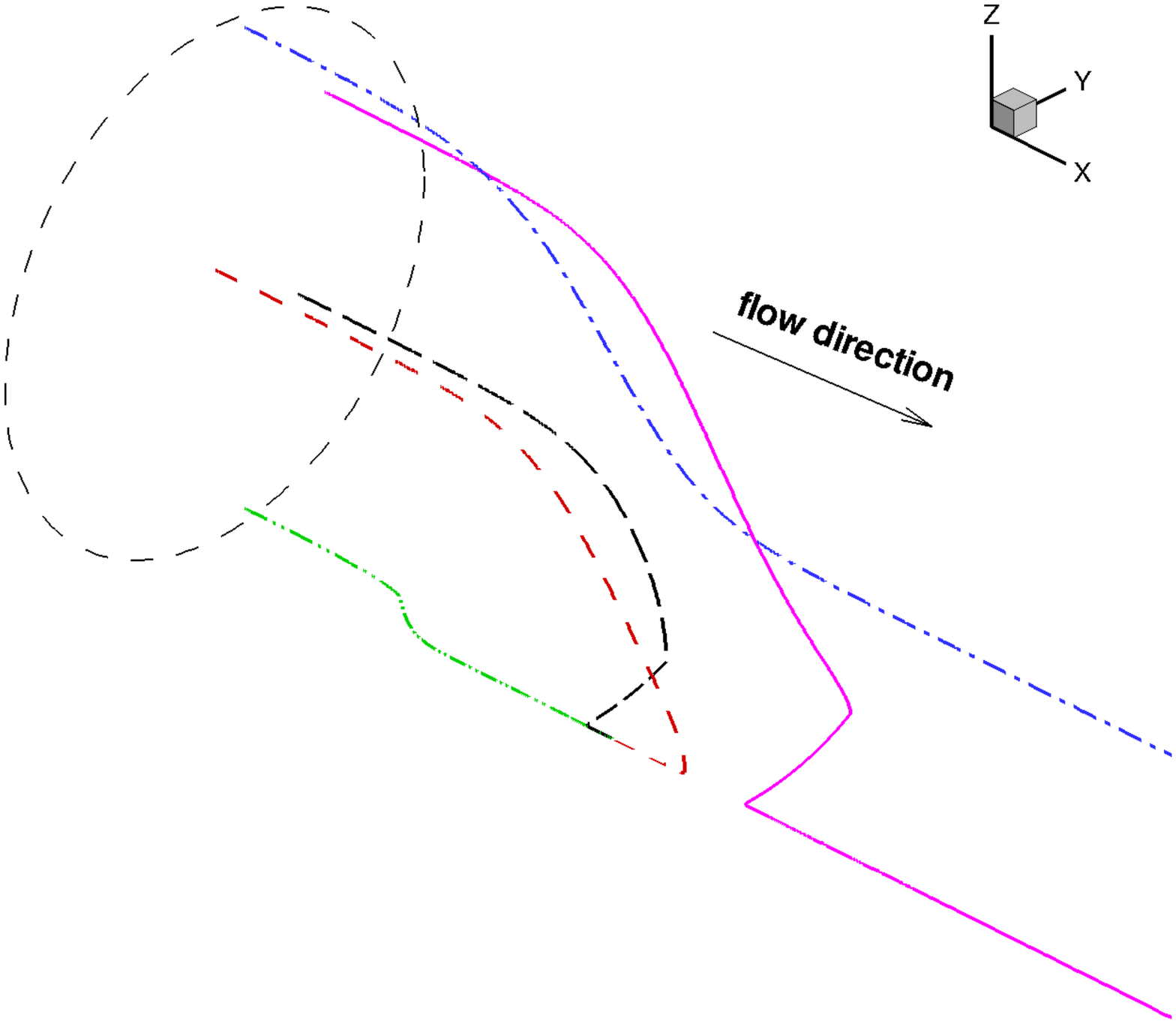}\hglue -0.3truein
\includegraphics[width=0.45\textwidth]{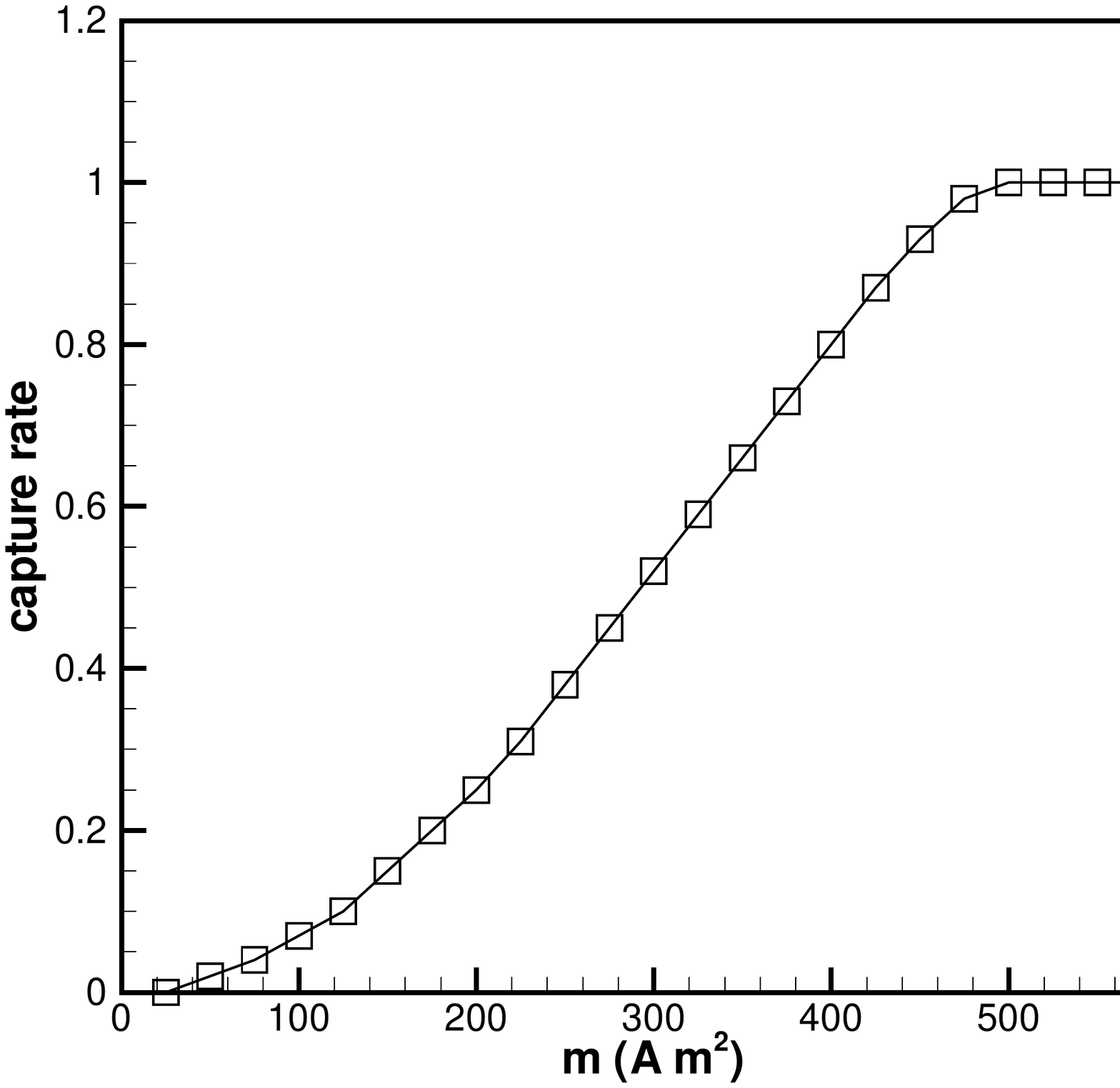}\hfil\\
\hfil (a)\hfil (b)\hfil\\
(c)\includegraphics[width=0.45\textwidth]{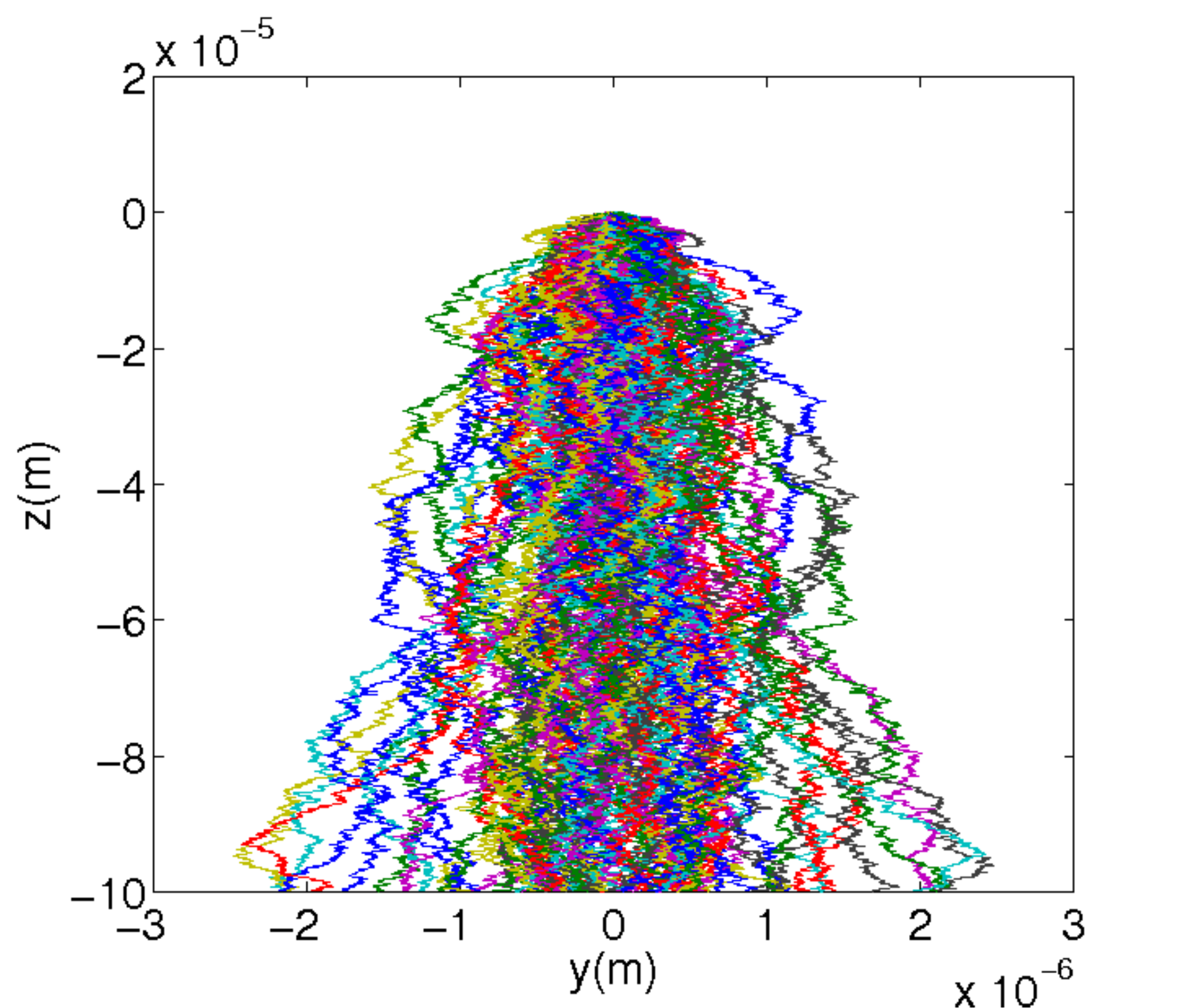}
(d)\includegraphics[width=0.45\textwidth]{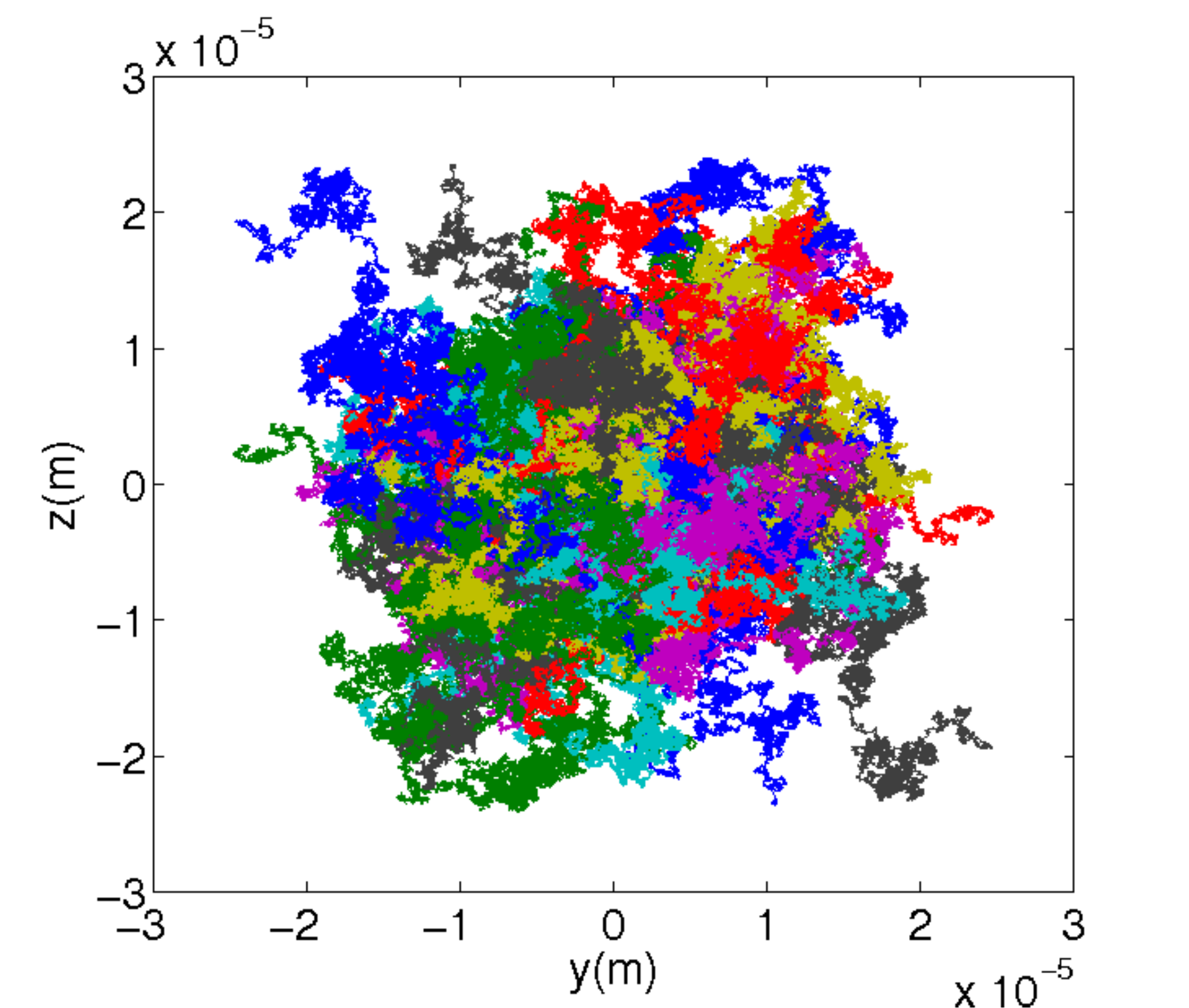}\hfil\\
\caption{Reproduced with permission from \cite{YLAR2012}. (a) Typical  trajectories with sliding motion on the wall. The scale is  compressed in the $x$-direction.   $R=10^{-4}$ m, $u_{max}=\frac{2}{\pi}\times 10^{-2}$ m s$^{-1}$,  $a = 10^{-7}$ m, $\eta=0.004$ Pa s, 
$\chi=0.2$, $m=1$A~m$^2$, and $d=50 R$. The clusters are released from $x=-100R$, the target is at $x=0$. 
(b) Capture rate {\it vs.} dipole moment $m$ with $d=0.03$~m, $\chi=0.2$.
(c) Trajectories in the $y-z$ plane with $x=0$ (target). $m=1056$~A~m$^2$, $\chi=0.2$, $\Delta t=2\times 
 10^{-4}$~s. 100 trajectories.
(d)  With smaller particles,  $a=10^{-8}$~m. }\label{fig:3dtraj}
\end{figure}

The system (\ref{system2})  is integrated using the Euler-Maruyama method. At the $n$th time step, where $\Delta t$ denotes a fixed step size, $X_{n+1}=X_n+\Delta t\ f(X_n)+g(X_n)\Delta W_n$.  As detailed in \cite{Higham2001},  the advantage of this method is the strong convergence under broad assumptions. We use the implementation 
in the SDE Toolbox for Matlab  \cite{Picchini}. 
The capture rate for a particle is defined to be the probability of hitting the tumor, which is situated  at $|x|<R_{tumor}$ at the wall in figure \ref{schematic}. 
Figure \ref{fig:3dtraj}(a) shows sample trajectories, initially at  $x=-R_{tumor}$, where $R_{tumor}$ denotes the radius of the tumor at the wall. A trajectory can escape without hitting the wall ($-.-$), or  escape after hitting the wall ($-$),  or be  captured at the target.  Figures   \ref{fig:3dtraj}(b-d) use data from \cite{ARRRP,House86}.  The magnet is $d=0.03$ m outside the vessel wall;  vessel radius $R=10^{-4}$~m, $u_{max}=\frac{2}{\pi} 0.01$~ms$^{-1}$, particle radius $a=10^{-7}$~m, $\eta= 0.004$Pa~s,  $R_{tumor}=100R$, constant susceptibility $\chi=0.2$.  
(b) shows the capture rate versus dipole moment $m$. For larger $m$, 
   a non-constant Langevin fit approximates the magnetization data  \cite{ARRRP}. The capture rate scales with the square of the dipole moment even when  the velocity profile is slightly blunted at the centerline which may relate to blood flow.  To optimize the capture rate, it is found that the distance $d$ should match the tumor size.   (c)  shows trajectories for $a=10^{-7}$~m, and compared with this, (d) shows that  Brownian motion makes a significant difference for    $a=O(10^{-8})$~m. 

The capture rate is optimal for a magnet that produces a strong field and a  focused effect, and therefore high field gradients. This is one of the technological challenges in magnetic drug targeting. Recent studies show that arrays of magnets, such as the Halbach array,  are an improvement over the single magnet. The balance of magnetic force and hydrodynamic drag for Stokes flow is calculated for high field gradient arrays and compared with experimental data in \cite{Barnsley2017}. This gives guidance on  how the particle trajectories depend on magnet shapes and arrays. 

\begin{figure}[t]
\begin{center}
\includegraphics[width=0.45\textwidth]{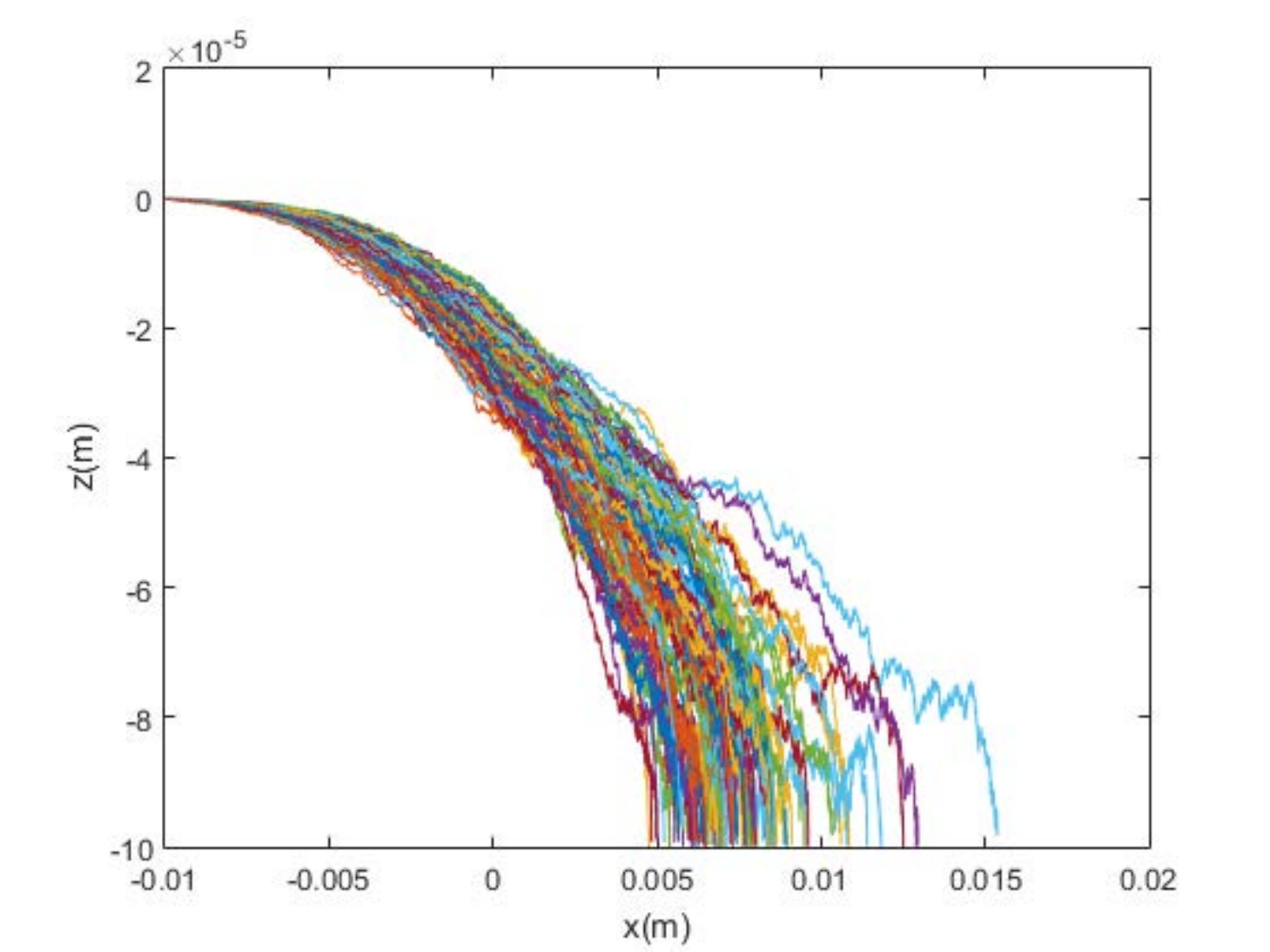}\hfil
\includegraphics[width=0.45\textwidth]{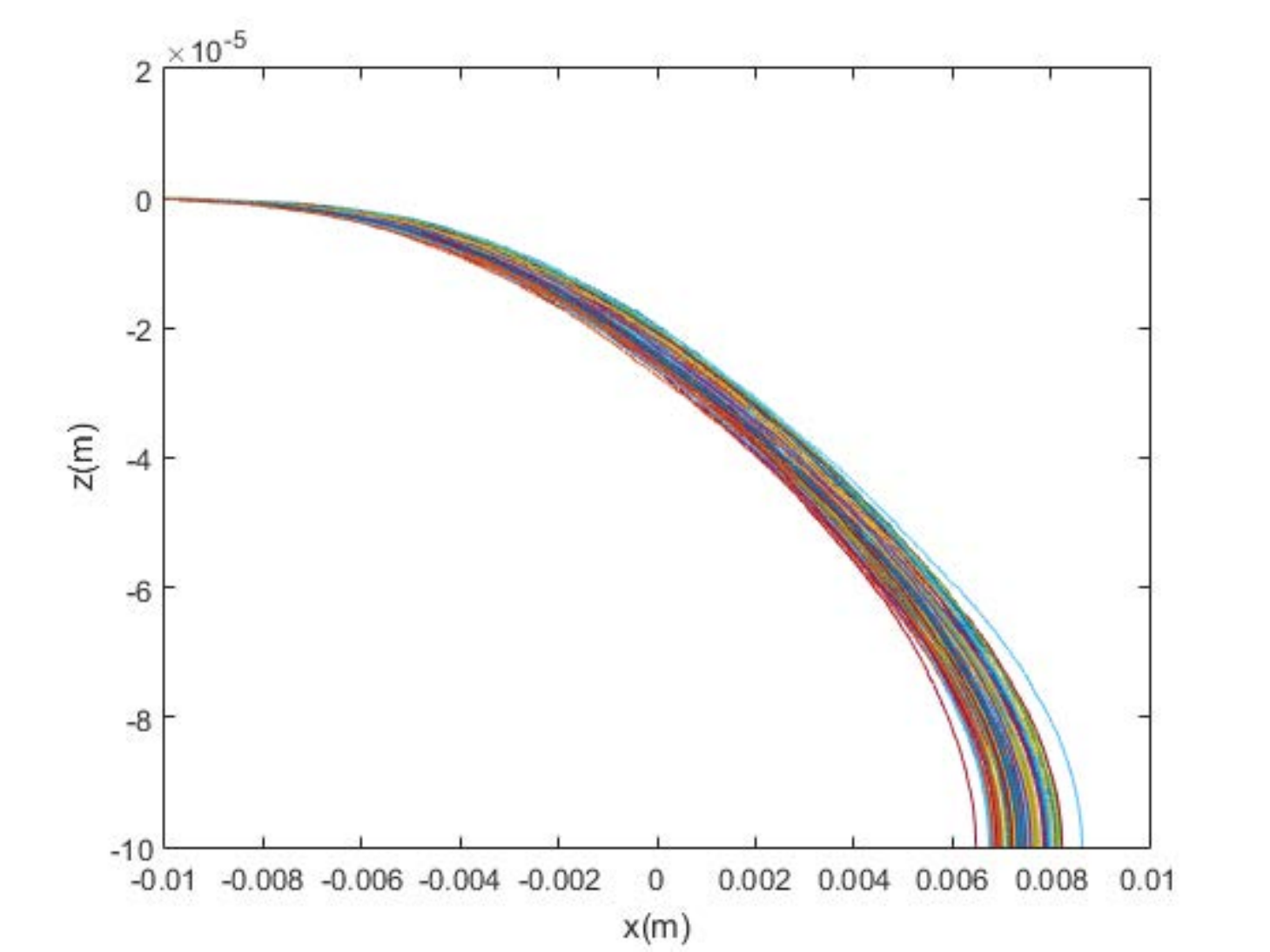}\\
\hfil (a) Z vs X with RBC \hfil (b) Z vs X without RBC\hfil \\
\end{center}
\caption{ (with permission from \cite{Mohren2017}, Figues 5(a-b)) 100 particle trajectories with point dipole magnetic field model and
Cherry viscosity model  with ${\bf m} =
[0, 0, 700Am^2]$. Without red blood cell collisions, the capture rate was 100\%. With red blood cell collisions, the capture rate was 91\%.}
\label{fig6}
\end{figure}

Non-Newtonian effects in blood flow have been studied with  empirical formulas that replace the Stokes drag coefficient \cite{Sprenger2015}. For example,  in the case of  the larger arterial speeds,  an empirical viscosity term which incorporates shear-thinning at the local shear rate   is applied to  write approximate equations for the motion of  micron sized magnetic particles in Poiseuille flow \cite{Cherry2014,Mohren2017}.  The particles in \cite{Cherry2014} are of micron size, much larger than the nanometers considered in \cite{YLAR2012}, and also the diameter of the artery and flow speeds are much larger,  and therefore the Brownian effect for the particles is small.   The use of the local shear rate of the background  flow in the Stokes drag formula is an ad-hoc substitution.   For a particle moving in non-Newtonian flow, such as in a shear-thinning fluid,  the calculation of drag is tractable for quiescent flow,  not for non-quiescent flow such as shear flow. Unlike Stokes flow, the equations are nonlinear so that  solutions can not be superposed.  The real velocity of the particle is spatially and temporally varying in response to the magnetic field and flow.  
 A separate effect  is  the influence of  red blood cells on the capture rate.  \cite{Mohren2017} applies a  formula for shear-enhanced diffusion in colloidal suspensions (equation (1) of \cite{Griffiths2012})  to estimate that the red blood cells contribute an
   additive term proportional to the shear rate inside the square root in equation \ref{system2}.  Particle trajectories for the Cherry viscosity model and  including the diffusion due to interaction with red blood cells are investigated in \cite{Mohren2017}. They find that the additional diffusion deteriorates the capture rate; figure \ref{fig6} shows this.

\FloatBarrier

%% file: NS2.tex
To investigate the response of a ferrofluid droplet to an applied
magnetic field, extra effort is needed to solve the details of the 
magnetic field, which in turn requires the knowledge of the
shape of the interface. The non-local moving boundary problem
which couples the spatial variation of the magnetic field with the 
interfacial shape of the drop necessitates a direct numerical computation. 
Numerical algorithms that explicitly track the interface and use finite
element methods are developed to investigate equilibrium shapes of
ferrofluid drops and interfacial instabilities in \cite{Lavrova,Lavrova04,Bashtovoi,Matthies,Knieling}. 
For the case of constant magnetic susceptibility, a boundary-integral 
formulation is used to study equilibrium shapes for high-frequency 
rotating magnetic fields \cite{Erdmanis2017}.  Numerical algorithms that 
implicitly track the interface include the Volume-of-Fluid (VOF) methods, 
Level-Set methods and diffuse-interface (DI) methods \cite{Scar99,Sethian2003,AMW}. 
Each of these methods can be applied to subcases of the results we show below. 
We specifically describe the VOF method because of its simplicity, efficiency and 
robustness for tracking topologically complex evolving interfaces. 
However, this as well as other approaches has its limitations. 
An advantage is that it conserves mass and with the recent improvements 
in the discretization of the surface tension force, it remains a competitive 
method for modeling interfacial flows. We present these results in section \ref{sec:vof}.

A recent study has shown that a ferrofluid droplet on a hydrophobic substrate 
breaks up into two daughter droplets and a smaller satellite droplet under a sufficiently
strong magnetic field \cite{Timonen2013}. These experiments also show that if the magnetic
field is increased even further, the droplets break up again and rearrange to form an assembly
of droplets on the substrate. In the same spirit, we present a model derived in \cite{Seric2014}
for the flow of a thin ferrofluid film on a substrate. The results in  \cite{Seric2014} show 
the film break up and the deformation of sessile droplets on the substrate under an external magnetic field.
We present this study in section \ref{sec:lub} 

\subsection{Governing equations}
\label{sec:vof}
The equations governing the motion of an incompressible ferrofluid drop 
suspended in another fluid are the conservation of mass and the balance of momentum
\begin{eqnarray}  
\frac{\partial \rho}{\partial t}+{\mathbf u}\cdot\nabla \rho&=&0,\label{eq:mass}\\
\rho \frac{\partial{\mathbf u}}{\partial t}+{\mathbf u}\cdot\nabla {\mathbf u}&=&-\nabla p + 
\nabla\cdot \bigl(2\eta {\mathbf D} \bigr) + \nabla\cdot{\mathbf \Pi}_M + \rho {\mathbf g},
\label{eq:motion}
\end{eqnarray}      
where  ${\mathbf u}({\mathbf x},t)$ is the velocity field, $p({\mathbf x},t)$ the pressure, 
$\rho({\mathbf x},t)$  and $\eta({\mathbf x},t)$ are the phase dependent density
and viscosity, respectively,
${\mathbf D}=\frac{1}{2}\big( \nabla{\mathbf u} +(\nabla{\mathbf u})^T\big)$ 
is the rate of deformation tensor (where $T$ denotes the transpose),
$\rho\mathbf{g} = - \rho g{\hat z}$ the body force due to gravity
and ${\hat z}$ the unit vector in the $z$-direction (with gravitational constant $g$),
and ${\mathbf \Pi}_M({\mathbf x},t)$ is the magnetic stress tensor
\begin{eqnarray} 
{\mathbf \Pi}_M  =  - \frac{\mu_0}{2} ({\mathbf H}\cdot{\mathbf H}){\mathbf I} +  \mu{\mathbf H}{\mathbf H},
\end{eqnarray}
where ${\mathbf I}$ is the identity tensor.
The Maxwell equations
for the  magnetic induction ${\mathbf B} $, magnetic field  ${\mathbf H}$, and  magnetization
${\mathbf M}$ are the magnetostatic Maxwell equations for a non-conducting ferrofluid 
\begin{eqnarray}
\nabla\cdot{\mathbf B}&=&0,\quad \nabla\times{\mathbf H}=0,\label{eq:main_eq}\\
{\mathbf B}({\mathbf x},t) &=&  \left\{ 
\begin{array}{ll}
                \mu_{d} {\mathbf H }  &  \mbox{in ferrofluid drop} \\
                \mu_m {\mathbf H}   &  \mbox{outside ferrofluid drop,}
\end{array}        
\right. 
\label{classical}
\end{eqnarray}
where $\mu_d$ denotes the permeability of the drop, and 
$\mu_m$ is the permeability of the surrounding fluid.  
Here, we assume that the surrounding fluid has a permeability
very close to that for a  vacuum, $\mu_0$. Therefore, we shall set $\mu_m=\mu_0$ 
throughout. A magnetic scalar potential $\psi$ is defined by
${\mathbf H}=\nabla\psi$, and satisfies
\begin{equation}
\nabla\cdot(\mu\nabla\psi)=0.  \label{eq:laplace}
\end{equation}  

For linear magnetic materials, the magnetization is a linear function of the magnetic field given 
by  ${\mathbf M} = \chi {\mathbf H}$,
where $\chi= (\mu/\mu_0 - 1)$ is the magnetic susceptibility. 
The magnetic induction ${\mathbf B}$ is therefore  
${\mathbf B}=\mu_0({\mathbf H}+{\mathbf M})=\mu_0(1+\chi){\mathbf H}$. 
To describe the paramagnetic susceptibility quantitatively, the Langevin function
 $L(\alpha)$=coth~$\alpha-\alpha^{-1}$  is used to describe the magnetization 
versus the magnetic field as 
\begin{eqnarray}
{\mathbf M} ({\mathbf H}) = M_s L\left(\frac{\mu_0 m |{\mathbf H}|}{k_B T}\right)\frac{{\mathbf H}}{|{\mathbf H}|}, \label{langevin1}
\end{eqnarray} 
where the saturation magnetization, $M_s$, and the magnetic moment of the particle, $m$,
enter as parameters, $T$ denotes the temperature, and $k_B$ is the Boltzmann's constant.

Let $\lbrack\!\lbrack \cdot \rbrack\!\rbrack$ denote the difference, $\cdot_{\rm surrounding}-\cdot_{\rm drop}$,  at the 
interface between the ferrofluid drop and the external liquid. Let ${\mathbf n}$ denote the unit normal 
outwards from the interface, and ${\mathbf t_1}$ and ${\mathbf t_2}$ denote the orthonormal tangent vectors. We 
require 
\begin{enumerate}
\item The continuity of velocity, the normal component of magnetic induction, the tangential component of 
the magnetic field, the tangential component of surface stress, 
\begin{eqnarray}
\lbrack\!\lbrack {\mathbf u}\rbrack\!\rbrack ={\mathbf 0},\quad
{\mathbf n}\cdot \lbrack\!\lbrack {\mathbf B}\rbrack\!\rbrack =0,\quad
{\mathbf n}\times \lbrack\!\lbrack {\mathbf H} \rbrack\!\rbrack ={\mathbf 0},\quad
 \lbrack\!\lbrack {\mathbf t_1}^T \cdot {\mathbf \Sigma}  \cdot {\mathbf n}  \rbrack\!\rbrack = 0,\quad
\lbrack\!\lbrack {\mathbf t_2}^T  \cdot {\mathbf \Sigma}   \cdot {\mathbf n} \rbrack\!\rbrack = 0.\nonumber
\end{eqnarray}

\item The jump in the normal component of stress balanced by capillary effects,
\begin{eqnarray}
\lbrack\!\lbrack {\mathbf n}^T \cdot {\mathbf \Sigma}  \cdot {\mathbf n} \rbrack\!\rbrack = 
\sigma k ,\nonumber
\end{eqnarray}
where ${\mathbf \Sigma}  = p - \rho g z + \bigl(2\eta {\mathbf D} \bigr) + {\mathbf \Pi}_M$,  
$k = -\nabla \cdot {\mathbf n}$ is the interface curvatures, and $\sigma$ 
is the interfacial tension.
\end{enumerate}


%% file: numerics2.tex
\subsection{One-fluid formulation and Volume-of-Fluid method}
\label{sec:vof}
We take advantage of the axial symmetry present in the 
motion and deformation of a ferrofluid droplet placed  in 
a viscous medium under an externally applied magnetic field 
and resort to the formulation in axisymmetric cylindrical coordinates $(r,z)$.
The  VOF method represents 
each liquid  with a color function as
\begin{eqnarray}
C(r,z,t) & = & \left\{
\begin{array}{ll}
0 & \mbox{in the surrounding medium}\cr
1 & \mbox{in the ferrofluid drop,}\cr
\end{array}
\right.
\end{eqnarray}
which is advected by the flow.  The position of the interface is 
reconstructed from 
$C(r,z,t)$.  The one-fluid formulation of the governing equations
(\ref{eq:motion}) and (\ref{eq:mass}) then reads
\begin{eqnarray}  
\frac{\partial C}{\partial t}+{\mathbf u}\cdot\nabla C&=&0, \label{eq:vof}\\
\rho \frac{\partial{\mathbf u}}{\partial t}+{\mathbf u}\cdot\nabla {\mathbf u}&=&-\nabla p + 
\nabla\cdot \bigl(2\eta {\mathbf D} \bigr) + \nabla\cdot{\mathbf \Pi}_M + {\mathbf F}_s + \rho {\mathbf g},
\label{eq:motion-vof}
\end{eqnarray}    
where
${\mathbf F}_s$ denotes the continuum body force due to interfacial tension \cite{Brackbill92}
\begin{equation}
{\mathbf F}_s=\sigma k  \delta_s {\bf n}, \label{forces}
\end{equation}
${\bf n} =\nabla C/|\nabla C|$, $\delta_s=|\nabla C|$ is the delta-function at the interface,   
$\rho = \rho(C)$, $\eta = \eta(C)$, and $\mu = \mu (C)$.
The magnetic potential $\psi$ in axisymmetric cylindrical coordinates satisfies equation
(\ref{eq:laplace}),
\begin{eqnarray}
\frac{1}{r}\frac{\partial}{\partial r}\left( \mu r\frac{\partial\psi}{\partial 
r}\right)+\frac 
{\partial}{\partial z} \left(\mu\frac{\partial \psi}{\partial z} \right)=0. 
\label{eq:phi}
\end{eqnarray}

The dimensionless variables, denoted by tildes, are defined as follows
\begin{align*}
  \tilde{\textbf{x}} = \frac{\textbf{x}}{L_c} 		&& \tilde{t} = \frac{t}{\tau}\\
  \tilde{\textbf{u}} = \frac{\tau}{L_c}\textbf{u}     &&\quad \tilde{\bf g} = \frac{{\bf g}}{g}\\
   \tilde{{\bf F}}_s = \frac{{\bf F}_s}{\sigma/L_c^2}            && \tilde{{\mathbf \Pi}}_M =  \frac{{\mathbf \Pi}_M}{H_o}\\ 
  \tilde{p} = \frac{p}{\sigma/L_c} && \tilde{\rho}=\frac{\rho}{\rho_d},\quad \tilde{\eta}=\frac{\eta}{\eta_d},\quad \tilde{\mu}=\frac{\mu}{\mu_0}
\end{align*}
where $L_c$, $\tau$, $H_o$ and $g$ are characteristic scales of length, time, magnetic field and gravitational acceleration, respectively. The ferrofluid density and viscosity are ${\rho_d}$ and ${\eta_d}$, respectively. For the choice of viscous time scale, 
$$
\tau = \frac{\eta_d L_c}{\sigma},
$$
corresponding to the viscous length scale
$$
L_{c} = \frac{\eta_d^2}{\rho_d\sigma},
$$
equation (\ref{eq:motion-vof}) becomes, dropping the tilde notation,
\begin{eqnarray}  
\mbox{Oh}^{-2}\left(\rho \frac{\partial{\mathbf u}}{\partial t}+{\mathbf u}\cdot\nabla {\mathbf u}\right)&=&-\nabla p + \nabla\cdot \bigl(2\eta {\bf D} \bigr) + \mbox{Bo}_{m}\nabla\cdot{\mathbf \Pi}_M + {\mathbf F}_s + \mbox{Bo} \rho{\mathbf g},
\label{eq:motion1}
\end{eqnarray}
where the Ohnesorge number $\mbox{Oh}=\eta_d/\sqrt{\rho_d \sigma L_c}$, the magnetic Bond number $\mbox{Bo}_{m}=\mu_0H^2_oL_c/\sigma$, and gravitational Bond number $\mbox{Bo}=\rho_d g L_c^2/\sigma$.

\subsection{Spatial discretization}\label{ap:discrete}
In axisymmetric coordinates $(r,z,\theta)$, the magnetic field
is given by
\begin{equation}
{\bf H}=(\frac{\partial\psi}{\partial r},\frac{\partial\psi}{\partial 
z},0),
\end{equation}
and the corresponding Maxwell stress is
\begin{equation}
{\mathbf \Pi}_M = \mu \left[\begin{array}{ccc}
\psi_r^2 - \frac{1}{2}\left(\psi_r^2 
+\psi_z^2\right)
&\psi_r\psi_z& 0 \\
\psi_r\psi_z& \psi_z^2 
-
\frac{1}{2}\left(\psi_r^2 +\psi_z^2\right) & 0 \\
0 & 0 & - \frac{1}{2}\left(\psi_r^2 + \psi_z^2\right) \end{array}
                \right]. \label{magnetic tensor}
\end{equation} 
We have
\begin{eqnarray}
\nabla\cdot{\mathbf \Pi}_M = \left[\frac{1}{r}\frac{\partial}{\partial r} [r  ({\Pi}_M)_{rr}]
\right. & + & \left. \frac{\partial}{\partial z} [ ({\Pi}_M)_{rz}] \right] \bf{e_r}  \\ \nonumber
& + &\left[\frac{1}{r}\frac{\partial}{\partial r} [r ({\Pi}_M)_{rz}] + \frac{\partial}{\partial z}
[ ({\Pi}_M)_{zz}] \right] \bf{e_z},
\end{eqnarray}
where $\bf{e_r}$ and $\bf{e_z}$ are unit vectors in $r$ and $z$ directions.
The magnetic potential field is discretized using second-order central differences.
In axisymmetric coordinates,
the discretization of  (\ref{eq:phi}) at cell $(i,j)$ yields, on a regular Cartesian grid of size $\Delta$, 
\begin{eqnarray}
\nabla\cdot(\mu\nabla\psi)_{i,j} &=&
\frac{1}{r_{i,j}}\frac{r_{i+1/2,j}\mu_{i+1/2,j}(\frac{\partial\psi}{\partial
r})_{i+1/2,j}
-r_{i-1/2,j}\mu_{i-1/2,j}(\frac{\partial\psi}{\partial r})_{i-1/2,j}}{\Delta}
\nonumber\\
&+&\frac{\mu_{i,j+1/2}(\frac{\partial\psi}{\partial
z})_{i,j+1/2}-\mu_{i,j-1/2}(\frac{\partial\psi}{\partial z})_{i,j-1/2}}{\Delta},
\end{eqnarray}
where, for instance for the cell face $(i+1/2,j)$,
\begin{eqnarray}
(\frac{\partial\psi}{\partial r})_{i+1/2,j} =
\frac{\psi_{i+1,j}-\psi_{i,j}}{\Delta}.
\end{eqnarray}

\begin{figure}
\begin{center}
\scalebox{0.5}{\includegraphics[trim=20mm 40mm 0 20mm, clip=true]{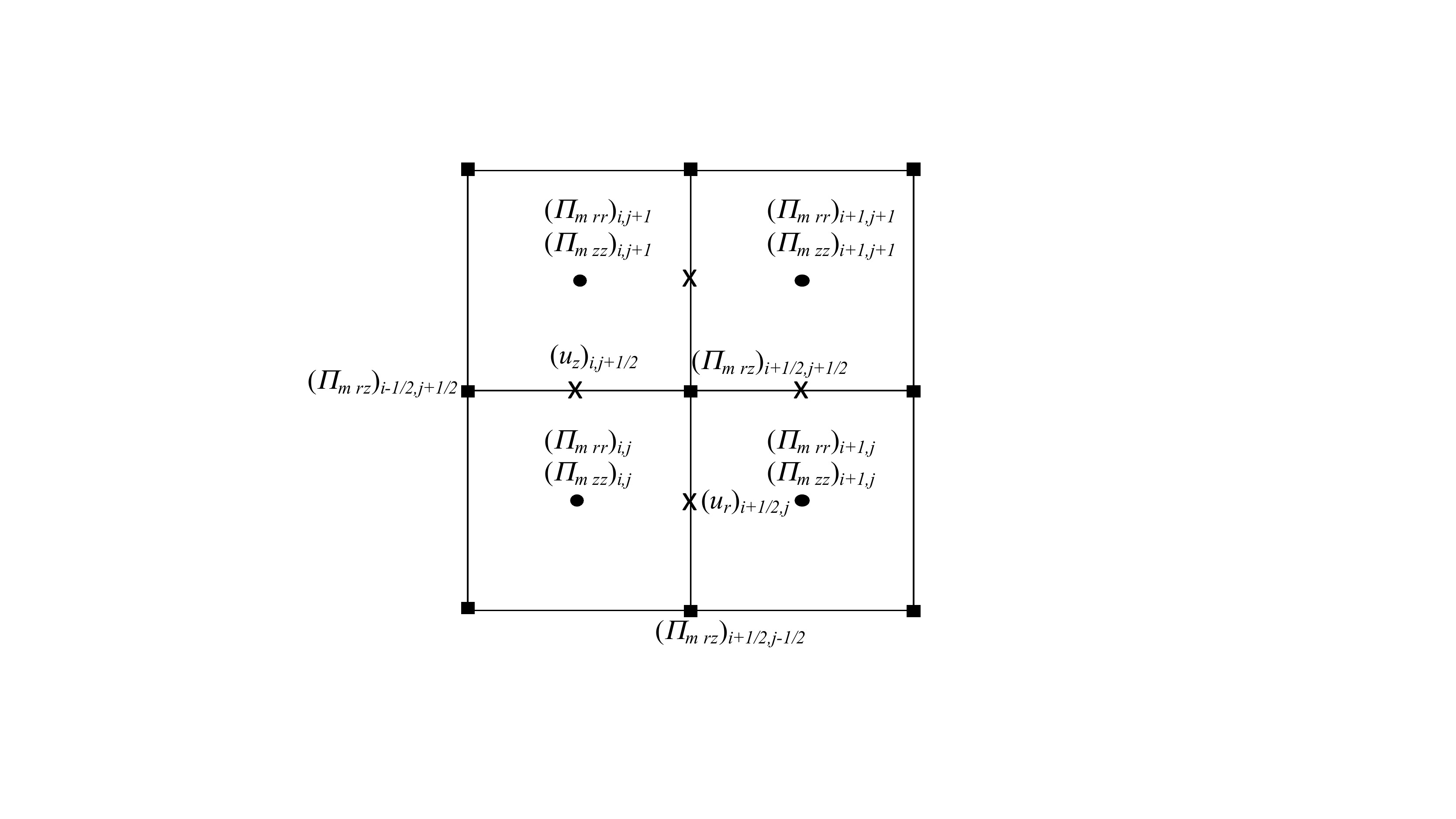}}
\end{center}
 \caption{Location of the velocities and the magnetic
stress tensor components on a
MAC grid.} \label{fig:figure22}
\end{figure}

The spatial discretization of the velocity field is based on the
MAC grid in figure \ref{fig:figure22}.
Therefore, the evaluation of the components of the magnetic stress tensor requires the evaluation
of gradients at faces. In axisymmetric coordinates,
the divergence of the magnetic stress tensor is discretized in the ${\bf e_r}$ and ${\bf e_z}$
directions as
\begin{eqnarray}
& & \frac{1}{r_{i+1/2,j}} \frac{r_{i+1,j}(({\Pi}_M)_{rr})_{i+1,j}
-r_{i,j}(({ \Pi}_M)_{rr})_{i,j}}{\Delta} + \nonumber\\
& &
\frac{(({\Pi}_M)_{rz})_{i+1/2,j+1/2}-(({ \Pi}_M)_{rz})_{i+1/2,j-1/2}}
{\Delta},\quad {\rm and} \nonumber\\
& & \frac{1}{r_{i,j+1/2}}
\frac{r_{i+1/2,j+1/2}(({ \Pi}_M)_{rz})_{i+1/2,j+1/2}
-r_{i-1/2,j+1/2}(({\Pi}_M)_{rz})_{i-1/2,j+1/2}}{\Delta} + \nonumber\\
& &
\frac{(({\Pi}_M)_{zz})_{i,j+1}-(({ \Pi}_M)_{zz})_{i,j}}{\Delta},
\end{eqnarray}
respectively, where, for example, components such as $(({\Pi}_M)_{rr})_{i,j}$ and
$(({ \Pi}_M)_{rz})_{i+1/2,j+1/2}$ are discretized as follows
\begin{eqnarray}
(({\Pi}_M)_{rr})_{i,j} &=& \mu_{i,j} \Bigl[\left(\frac{\psi_{i+1,j} - \psi_{i-1,j}}{2\Delta}\right)^2 \nonumber\\
&-& \frac{1}{2}\left(\left(\frac{\psi_{i+1,j} - \psi_{i-1,j}}{2\Delta}\right)^2 + \left(\frac{\psi_{i,j+1} -
\psi_{i,j-1}}{2\Delta}\right)^2\right) \Bigr], \nonumber
\end{eqnarray}
\begin{eqnarray}
(({\Pi}_M)_{rz})_{i+1/2,j+1/2} = \mu_{i+1/2,j+1/2}
\left(\frac{\psi_{i,j} - \psi_{i-1,j} + \psi_{i,j-1} - \psi_{i-1,j-1}}{2\Delta}\right)
\times \nonumber \\
\left(\frac{\psi_{i-1,j} - \psi_{i-1,j-1} + \psi_{i,j} - \psi_{i,j-1}}{2\Delta}\right),
\end{eqnarray}
respectively, where $\mu_{i+1/2,j+1/2}$ is computed using a simple averaging from cell
center values. $({\Pi}_M)_{rr}$ and $({ \Pi}_M)_{rz}$ at other grid locations
are discretized similarly. Analogous relationships can be written for the other components
of the magnetic stress tensor. 

\subsection{Drop deformation under a uniform magnetic field} 
\begin{figure}
\begin{center}
\includegraphics[scale=0.4]{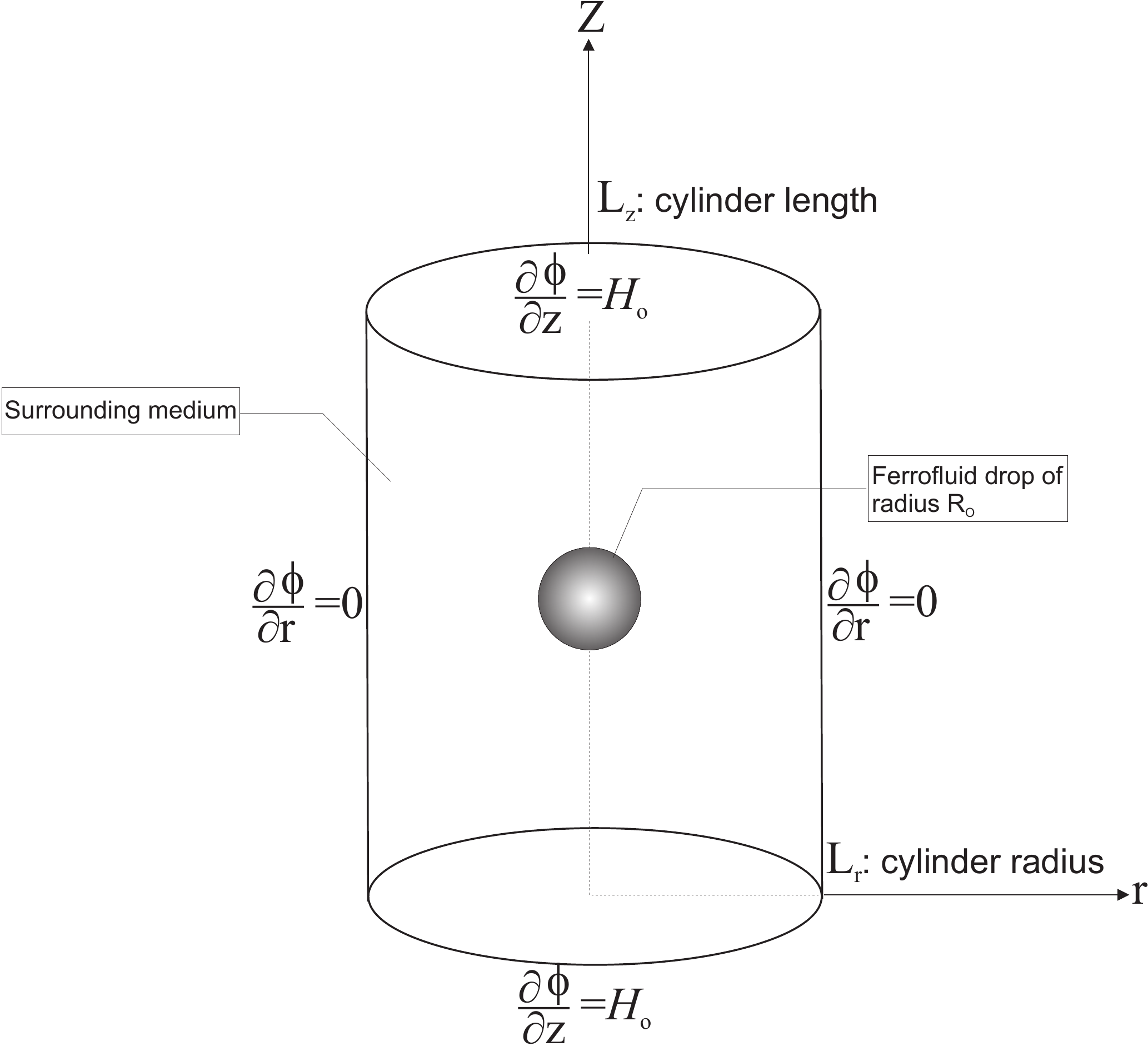}
\caption{Schematic of the initial configuration. The computational domain is $0\leq z\leq 
L_z$,  $0\leq r \leq L_r$. Initially, a spherical superparamagnetic ferrofluid drop of 
radius $R$ is 
centered in the domain. The boundary conditions on the magnetic field are depicted at the 
boundaries. Reproduced from \cite{ATRRWPR}.} 
\label{fig:figure4}
\end{center}
\end{figure}

When a ferrofluid drop is suspended in a non-magnetizable medium in an externally applied 
uniform magnetic field, it elongates in the direction of the applied magnetic field and assumes
a stable equilibrium configuration achieved via the competition  between the capillary and magnetic
forces.  Next, we show the numerical results of the VOF method for a ferrofluid drop suspended  in a non-magnetizable viscous medium by Afkhami et al.~\cite{ATRRWPR}. 
The initial configuration for the computational study of a ferrofluid drop  suspended 
in a non-magnetizable viscous medium is shown in figure~\ref{fig:figure4}.
A uniform magnetic field  ${\bf H}=(0,0,H_o)$, where $H_o$     
is the magnetic field intensity at infinity, is imposed at the top and  bottom boundaries
of the computational domain. In order to solve  Laplace's equation (\ref{eq:phi}) in the presence 
of an interface, the 
following boundary conditions are employed:
${\frac{\partial}{\partial z}\psi} = H_o$ at  $z=0, L_z$, and
${\frac{\partial}{\partial r}\psi} = 0$ at the side boundary $r=L_r$.
Note that a symmetry condition, ${\frac{\partial}{\partial z}\psi} = 0$, at $z = 0$,
and ${\frac{\partial}{\partial r}\psi} = 0$, at $r = 0$ can be applied. 
The drop radius is $R_o = 1$~mm,  
interfacial tension is $\sigma = 1$ mN~m$^{-1}$, and 
$\mu_d$ is chosen to be constant. 

\begin{figure}
\begin{center}
(a)\includegraphics[scale=0.4]{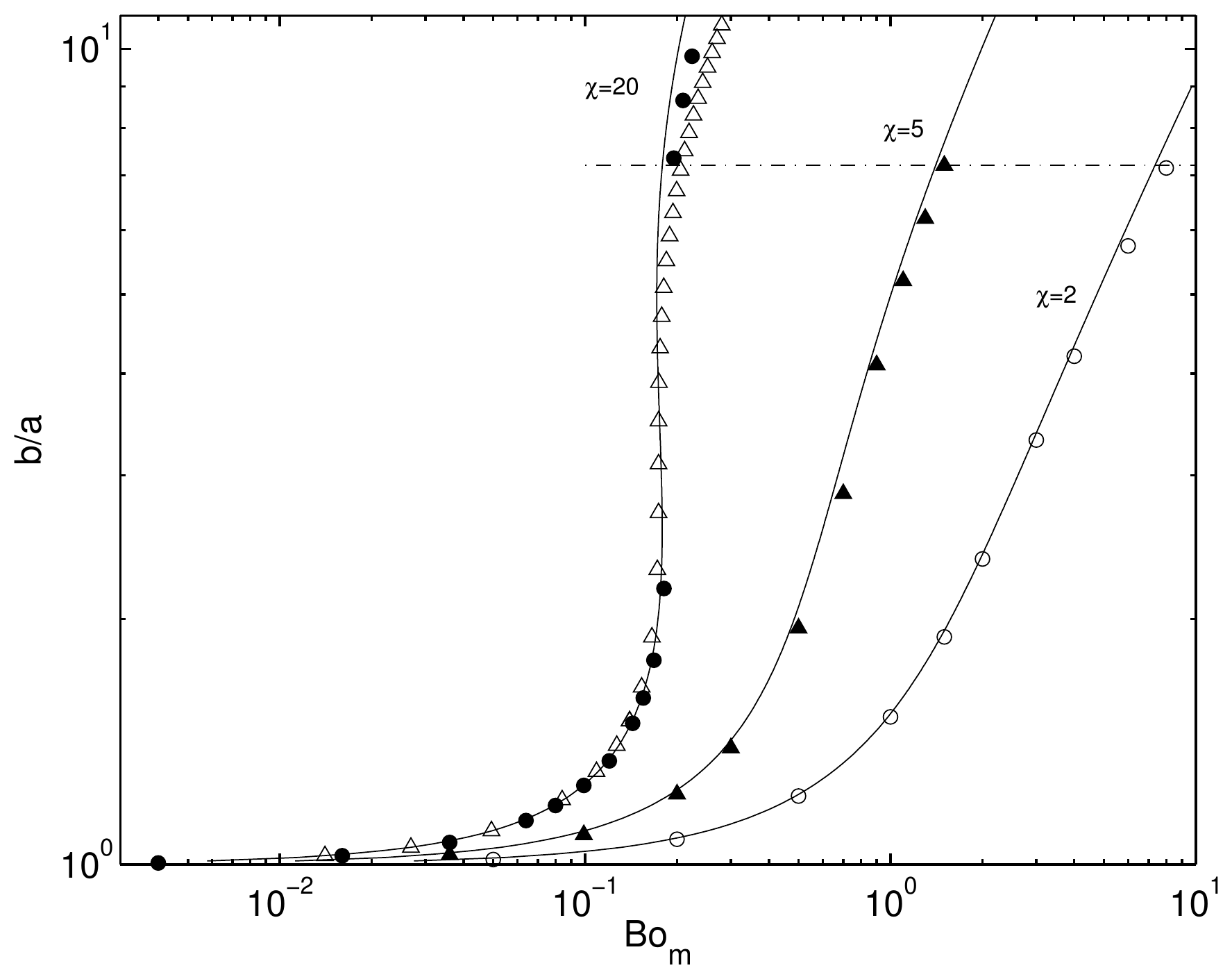}
(b)\includegraphics[height=59mm, trim = 25mm 1.5mm 72mm 2.5mm, clip=true]{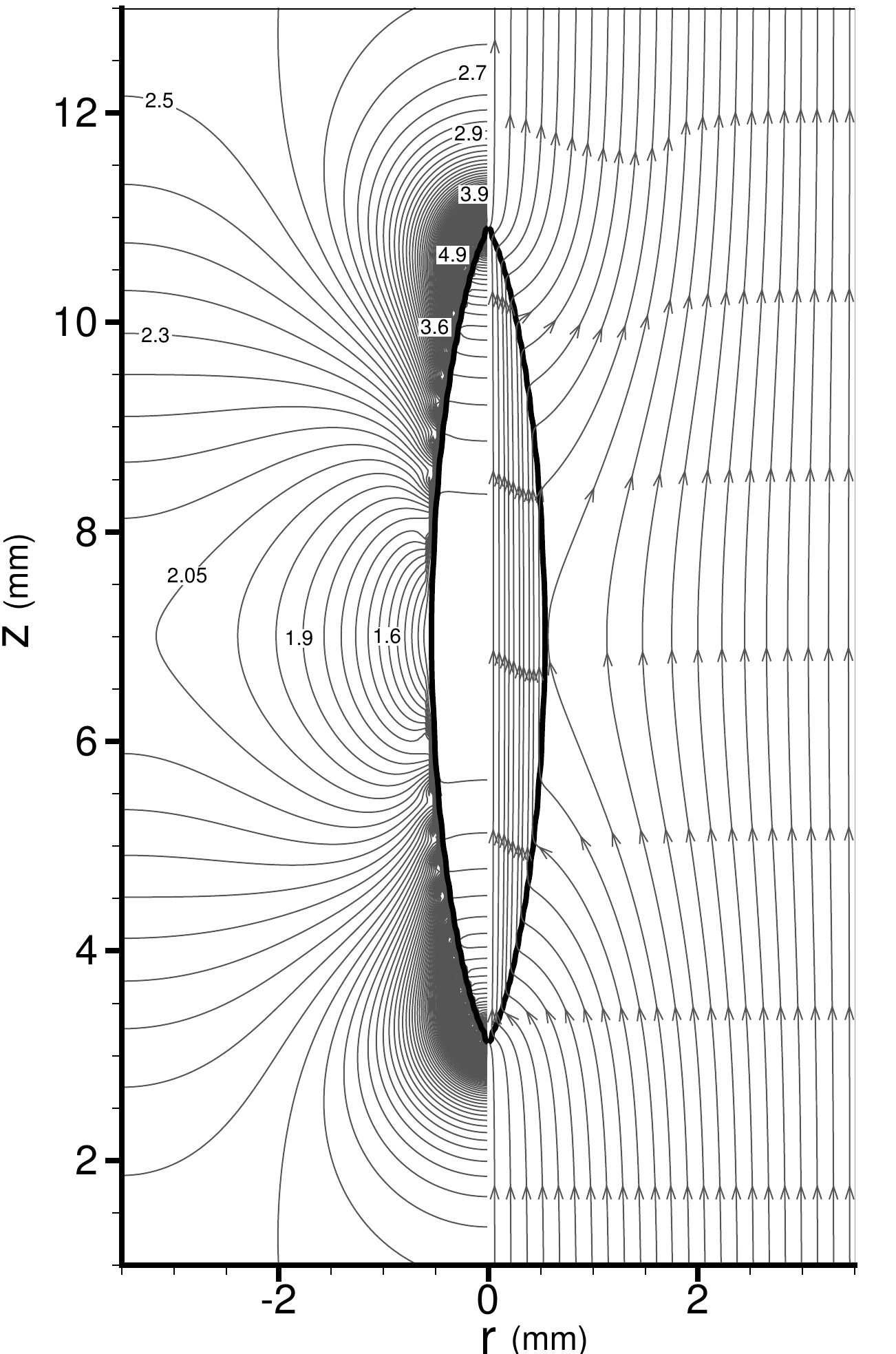}
\caption{From \cite{ATRRWPR}. (a) Comparison of the dependence of the drop aspect ratio $b/a$
(where $b$ is the semi-major axis and $a$ the
semi-minor axis) on the magnetic Bond number $\mbox{Bo}_{m}$.
Solid lines represent the theoretical analysis in \cite{ATRRWPR}. Numerical 
results are presented for $\chi = 2$ ({\large $\circ$}), $\chi = 5$ ($\blacktriangle$) and $\chi = 20$ ({\large $\bullet$}). 
(\small $\triangle$) denotes the prediction in \cite{Bacri}.
(b) The drop shape and contours of the magnetic field amplitude and magnetic field lines (right)
for the highly deformed drop suspended in a non-magnetizable media corresponding to $b/a\approx7$,
$\chi = 20$, and $\mbox{Bo}_{m} = 0.2$ data point in (a).}
\label{fig:figure7}
\end{center}
\end{figure}

Figure \ref{fig:figure7}(a) shows the results in \cite{ATRRWPR} along with theoretical predictions.
The details of the theoretical analysis is given in \cite{ATRRWPR}. 
 From the computational results alone, it is observed that at sufficiently large 
 $\chi$, the drop equilibrium aspect ratio jumps to a higher value when the 
 magnetic Bond number reaches a critical value, while 
for small values of $\chi$, the drop equilibrium aspect ratio increases continuously as a function of the magnetic Bond number, in agreement with the theories. Specifically, at $\chi=20$, the jump in drop shape  occurs 
 when the magnetic Bond number changes from $\mbox{Bo}_{m} = 0.18$ to $0.19$.
they also find that for a fixed 
Figure \ref{fig:figure7}(b) shows the result for $b/a \approx 7$, $\mbox{Bo}_{m} = 0.2$, and $\chi = 20$
(see the line -.- in part a), where the shape exhibits 
conical ends. To provide further insight, 
the magnetic field lines inside the highly deformed drop as well as in the 
non-magnetizable surrounding medium, showing the effect of the magnetic field responsible
for the appearance of conical ends. 

\subsubsection{Drop deformation under non-uniform magnetic fields} 
\begin{figure}
\begin{center}
\includegraphics[scale=0.45,trim = 50mm 0mm 50mm 0mm, clip=true,angle=-90]{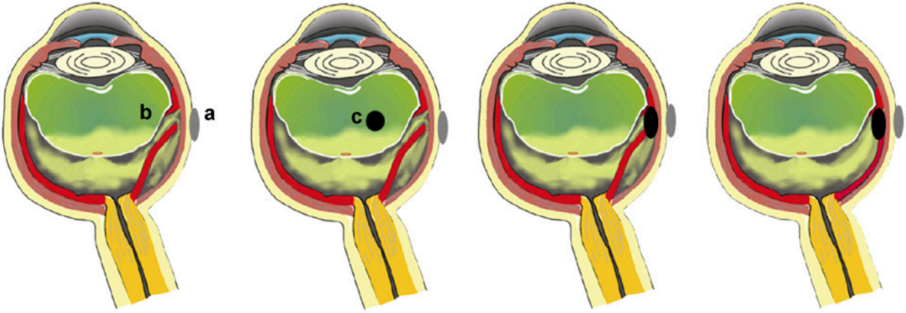}
 \caption{Schematic of a procedure for treating the retinal detachment. An external permanent magnet (a),
chosen based on its suitability for the given application to the eye surgery,
 is placed on the eye near the detachment site (b). \
 A ferrofluid droplet (c) is then injected into the eye. The external magnet guides the drop
to the site of the tear to seal it. Reproduced from \cite{Mefford}. 
} \label{fig:retina}
\end{center}
\end{figure}

Here we present the motion of a hydrophobic ferrofluid droplet placed  in 
a viscous medium and driven by an externally applied non-uniform magnetic field is 
investigated numerically in an axisymmetric geometry. 
This numerical investigation is motivated by recent developments in the
synthesis and characterization of ferrofluids for possible use in
the treatment of 
retinal detachment \cite{Mefford}. 
Figure \ref{fig:retina} shows a cartoon of the application of 
a small ferrofluid drop injected into the vitreous cavity of the eye and guided by a permanent  magnet 
inserted outside the scleral wall of the eye.
The drop travels toward the side of  the eye, until it can seal a retinal hole. 

\begin{figure}
\begin{center}
\includegraphics[scale=0.5,trim = 0mm 80mm 40mm -20mm, clip=true]{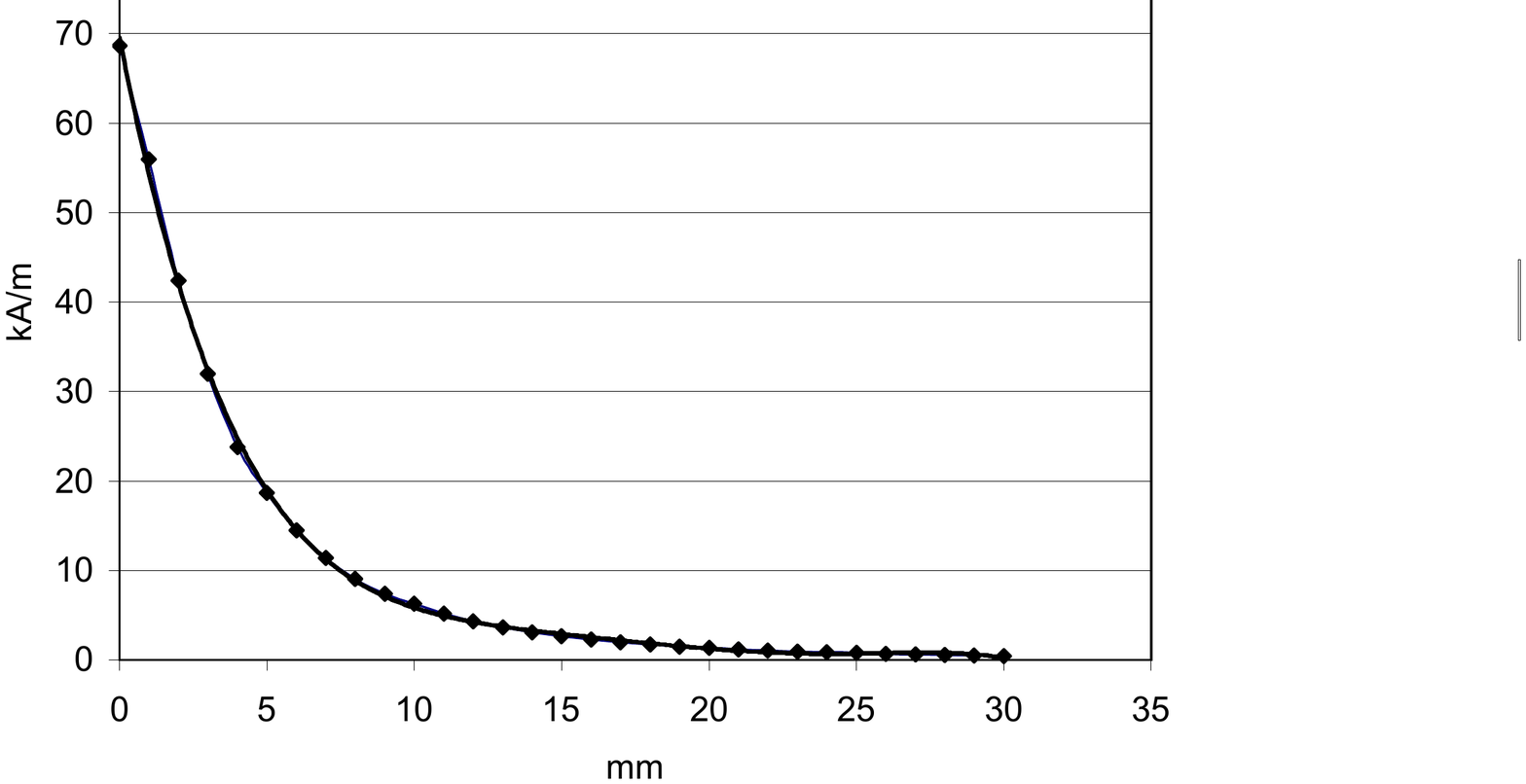}
 \caption{From \cite{ARRRP}. Measured data for the magnetic field from \cite{Mefford} 
($\blacklozenge$) and 
fifth degree  polynomial  fitted to the data (---) as functions of the distance from the 
magnet.}\label{fig:polynomial}
\end{center}
\end{figure}

To better understand the motion of the ferrofluid droplet
moving in the eye, Afkhami et al.~\cite{ARRRP} present 
a numerical investigation of a more straightforward scenario
where an initially spherical drop is placed at a height $L$ above 
the bottom of the cylindrical domain depicted in figure \ref{fig:figure4}. 
In addition, the boundary condition in the figure is changed to reflect 
the presence of a magnet at the bottom, which instantly magnetizes the drop. 
The boundary condition on the magnetic field is reconstructed from the experimental 
measurements in 
\cite{Mefford}, where 
the magnitude of $H(z)$ is measured as a function of distance from 
the magnet, $z$, in the absence of the drop.  
We then fit the data to a 5th degree polynomial, as shown in  
figure~\ref{fig:polynomial}.  
The scalar potential therefore is 
a 6th degree polynomial $\phi(0,z)=P_6(z)$ along the axis of the cylindrical domain.
In the absence of the drop, $\psi$ satisfies Laplace's equation 
$\frac{1}{r}\frac{\partial}{\partial r}(r\frac{\partial\psi}{\partial r})+\frac 
{\partial^2 \phi} {\partial z^2}=0$. If there is a solution, it is analytic and has 
$r^2$-symmetry.  The ansatz $\phi(r,z)=P_6(z)+r^2P_4(z)+r^4P_2(z)+r^6P_0(z)$ yields
\begin{eqnarray}
\phi(r,z)=P_6(z)-\frac{1}{4}r^2P_6''(z)+\frac{1}{64}r^4P_6^{(iv)}(z)-\frac{1}{(36)(64)}r^6P_6^{(vi)}(z). \label{phi}
\end{eqnarray}
This yields the boundary condition, and also approximates an initial condition 
when the drop is relatively small.  
The magnetic potential $\psi$ is calculated from equation (\ref{eq:phi}).
The boundary conditions for $\psi$ 
on the domain boundaries $\partial \Omega$ are defined as
\begin{eqnarray}
\frac{\partial \psi}{\partial n_b} = \frac{\partial \phi}{\partial n_b}, 
\end{eqnarray}
where $\partial /\partial n_b={\bf n}_b\cdot\nabla$, and ${\bf n}_b$  denotes the normal to the 
boundary $\partial \Omega$.
In order to impose the boundary condition in the numerical model,
a transformation of variables is performed to $\zeta$: $\phi=\psi+\zeta$, where
$\phi$ is the potential field without the magnetic medium.
One can then rewrite equation (\ref{eq:phi}) such that
\begin{equation}
\nabla\cdot(\mu\nabla\zeta)=-\nabla\cdot(\mu\nabla\phi),\label{magnet2}
\end{equation}
where $\nabla\cdot(\mu\nabla\phi)$ vanishes everywhere except on the surface
between the drop and the surrounding fluid $\partial \Omega_f$ and 
\begin{eqnarray}
\frac{\partial \zeta}{\partial n} = 0 & \mbox{on $\partial \Omega$}.\label{BCmagnet2}
\end{eqnarray}

\begin{figure}
\begin{center}
\includegraphics[trim = 70mm 25mm 70mm 40mm, clip, height=90mm]{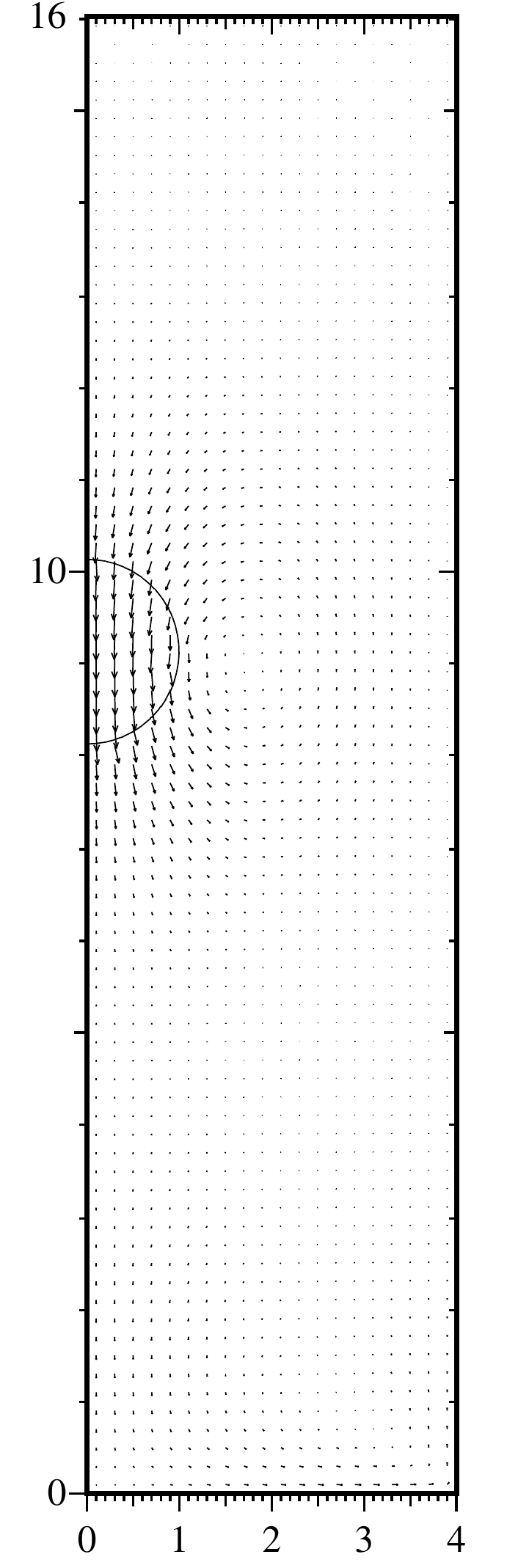}
\includegraphics[trim = 70mm 25mm 70mm 40mm, clip, height=90mm]{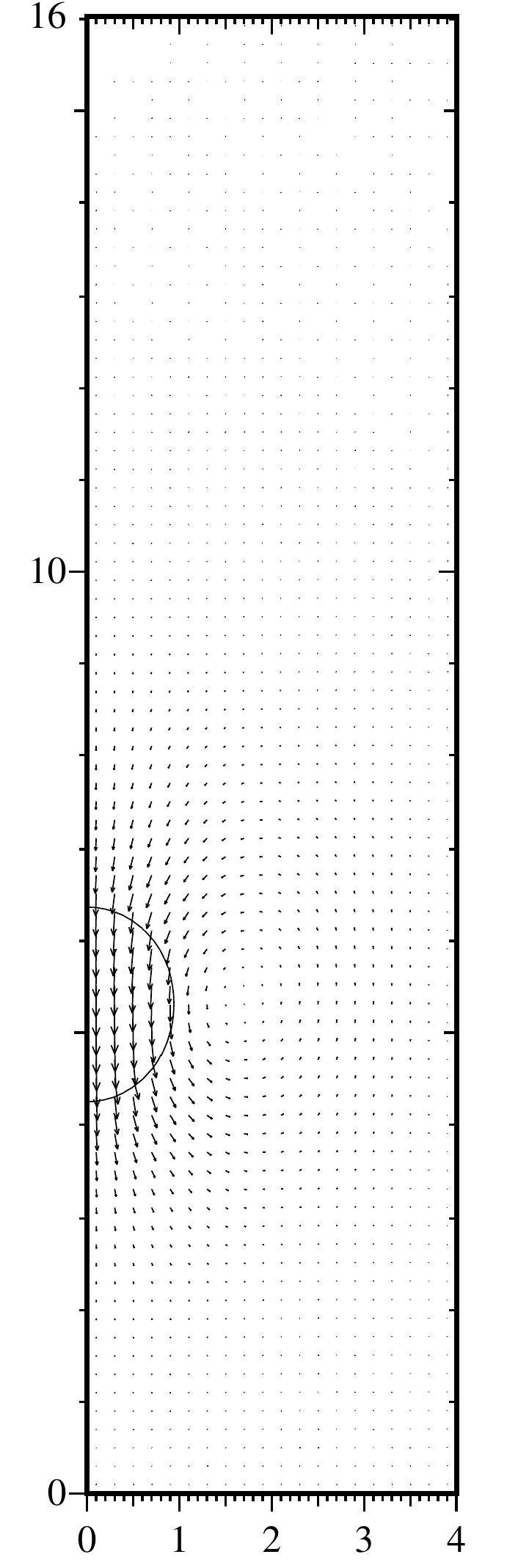}
\includegraphics[trim = 70mm 25mm 70mm 40mm, clip, height=90mm]{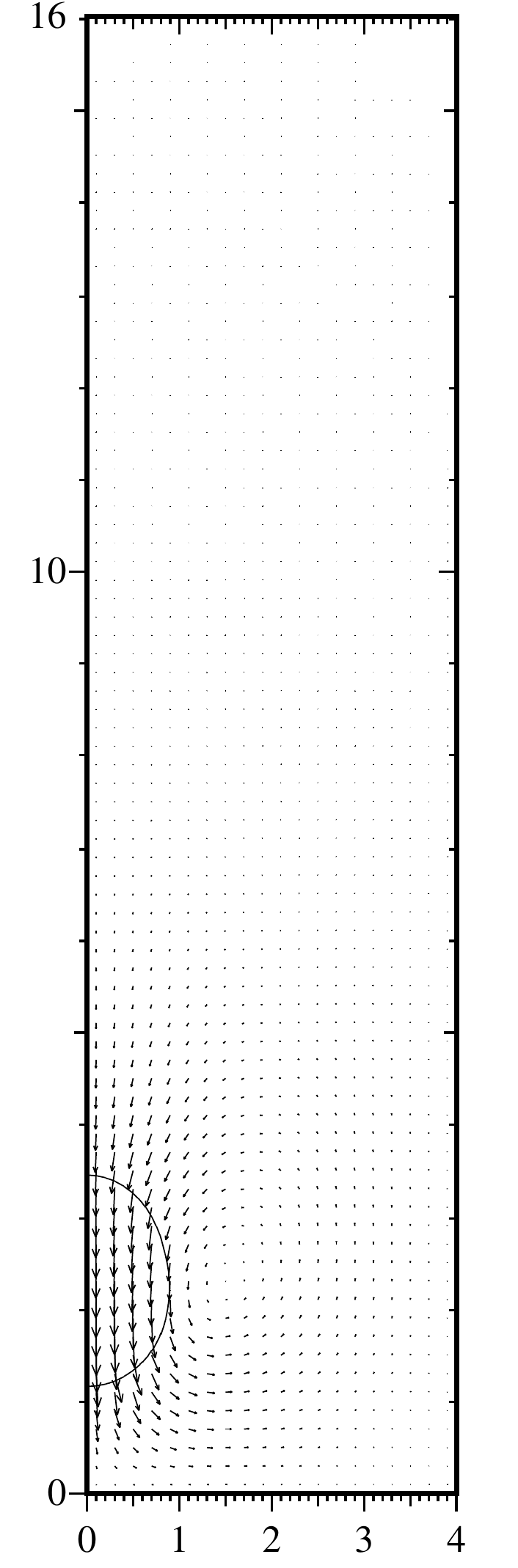}
\includegraphics[trim = 70mm 25mm 70mm 40mm, clip, height=90mm]{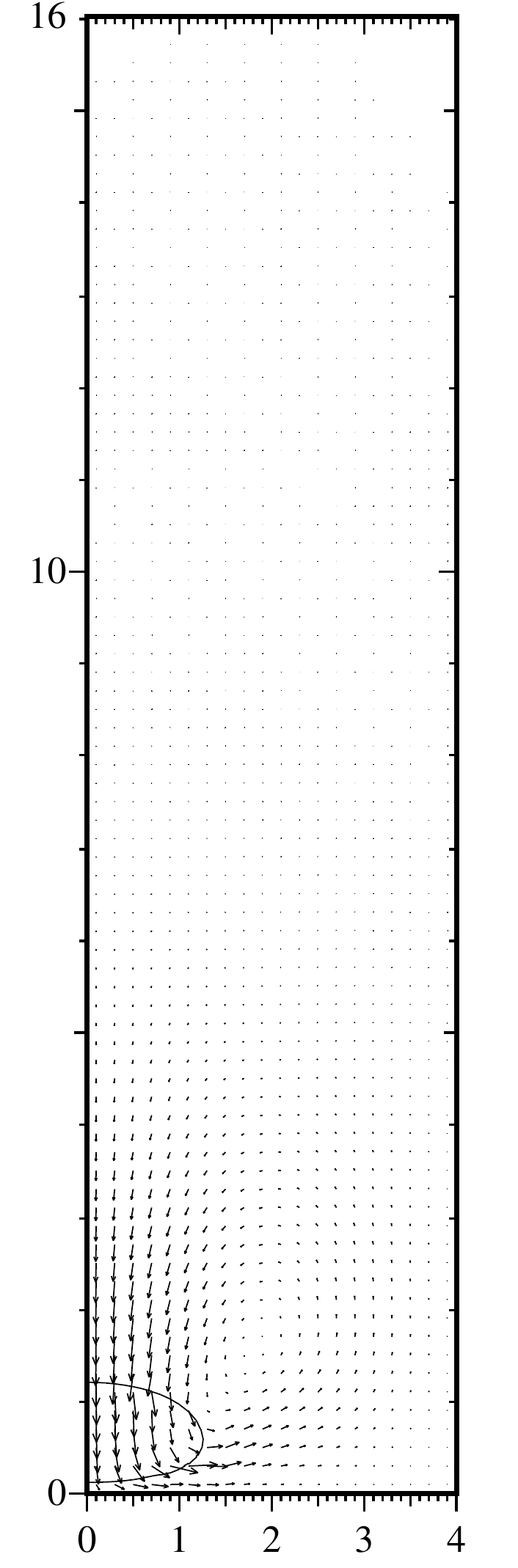}
\end{center}
\caption{Drop shapes along with the velocity fields at times $t=120, 160$, and 170~s (from left to right) 
for a droplet of 2~mm diameter placed 11~mm away from the bottom of the domain,
where the magnet is placed. For these simulations, $\mbox{Bo}_m = 0.06$. Reproduced from \cite{ARRRP}.} \label{fig:droplet_velocity}
\end{figure}

Figure~\ref{fig:droplet_velocity} shows direct numerical simulation results 
for one of the series of experiments conducted in 
\cite{Mefford}: a droplet of 2~mm diameter placed 11~mm away
from the bottom of the domain where the permanent magnet is placed.
The figure shows the migration of the drop at times
t~=~120, 160, and 170~s  along with the velocity fields. 
The travel time compares well with the experimentally measured one in \cite{Mefford}.
Additionally, the results show that at an early stage, the flow occurs approximately downwards and 
only in the region close to the droplet. As the droplet approaches the magnet,
it elongates in the vertical direction and vortices induced in the viscous medium become stronger.
When the droplet reaches the bottom of the domain, the flow inside the droplet is pumped
outward from the center of the droplet, resulting in the flattening of the droplet
and consequently a decrease in the droplet height. These observations are consistent
with experiments in \cite{Mefford}.


\subsection{Magnetowetting of thin ferrofluid films}
\label{sec:lub}
\begin{figure}
\begin{center}
\includegraphics[scale=0.475,trim = 70mm 80mm 60mm 90mm]{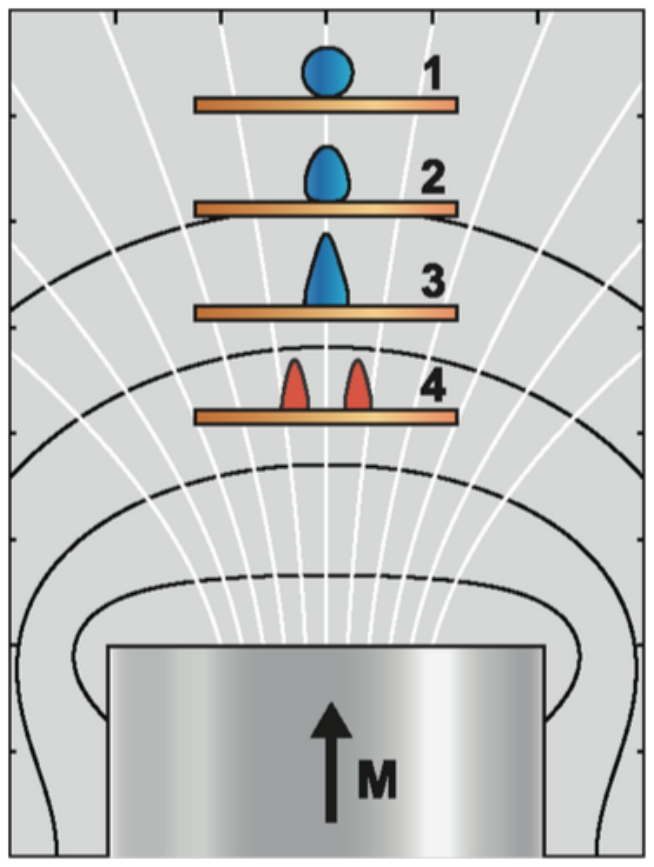}
\includegraphics[scale=0.475,trim = 20mm 80mm 30mm 150mm]{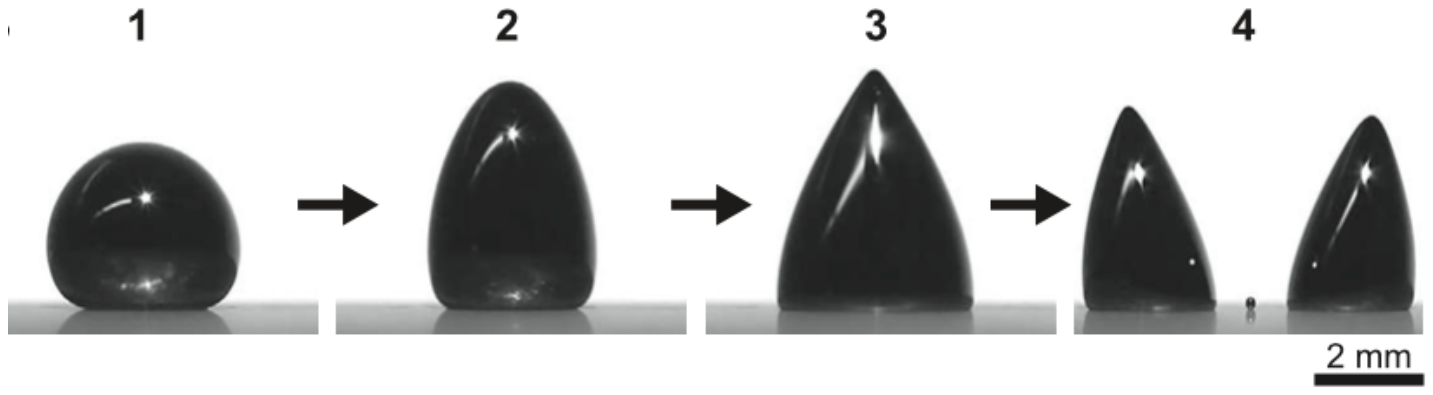}
 \caption{Schematic side-view of the
experiments from \cite{Timonen2013}. State
of the droplet is: 1, near-zero field; 2,
weak field; 3, strong field; and
4, above critical field (drop splits to two daughter droplets). (B) Experimental photographs from \cite{Timonen2013} of a 20-ml ferrofluid droplet upon increasing the field
from 80 Oe (dH/dz 3.5 Oe/mm) to 680 Oe (dH/dz 66 Oe/mm).
}\label{fig:science}
\end{center}
\end{figure}

Figure \ref{fig:science} shows the phenomenon of magnetowetting
for an experiment with magnetic droplets on a superhydrophobic surface, below 
which is a permanent magnet \cite{Timonen2013}. By gradually increasing the strength of the magnetic
field and the vertical field gradient, the figure shows the transition of the the droplet from a 
spherical shape into a spiked cone and eventually splitting
into two smaller droplets at a critical field strength. 

In \cite{Seric2014}, a problem closely related to  magnetowetting is investigated:  
the application of a uniform magnetic field to induce dewetting of a thin ferrofluid film. 
Here we present the study in \cite{Seric2014}, where
thin film equations are derived using the long wave approximation of the 
coupled static Maxwell and Stokes equations and 
the contact angle is imposed via a disjoining/conjoining pressure model. 

\begin{figure}
  \centerline{\includegraphics{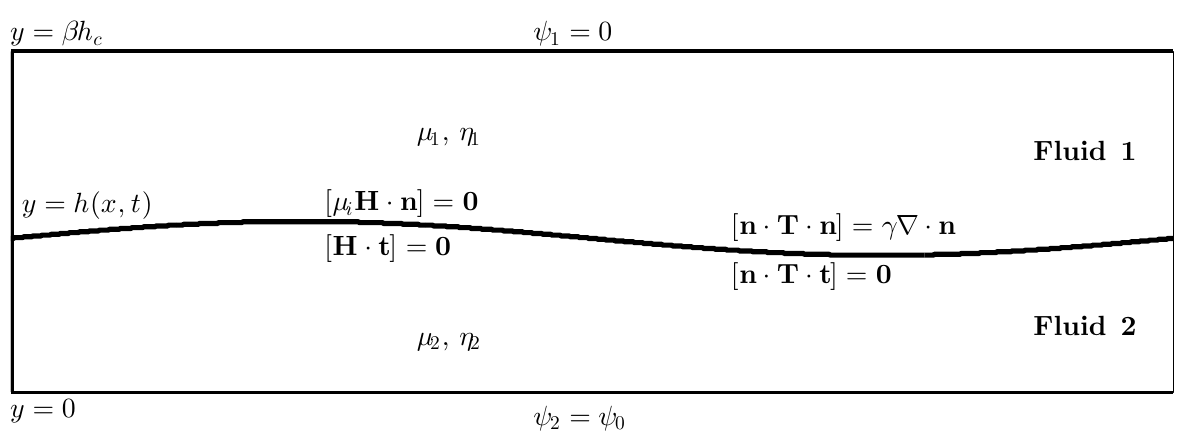}}
  \caption{From \cite{Seric2014}. The schematic of the system of two thin fluids, where fluid $1$ is nonmagnetic and fluid $2$ is ferrofluid.}
\label{fig:lub}
\end{figure}

Figure \ref{fig:lub} shows the schematic of the system of two thin fluid
films in the region $ 0 < y < \beta h_c $, with the ferrofluid film occupying the region $ 0 < y < h_c $, 
and the nonmagnetic fluid occupying the rest of the domain.  We denote the permeability and viscosity of the 
fluids by $\mu_i$ and $\eta_i$, where $i=(1,2)$ denotes fluids $1$ and $2$, respectively. The interface between
two fluids is denoted by $y=h(x, t)$.   
The unit normal and tangential vectors to the interface, $\textbf{n}$ and $\textbf{t}$, respectively, are given by
\begin{equation}
\textbf{n} = \frac{1}{\left(1+h_x^2\right)^{1/2}}\left( - h_x, 1\right) ,\,\,\,\,\,\,\,\,\,\, \textbf{t} = \frac{1}{\left(1+h_x^2\right)^{1/2}}\left(1, h_x\right).   \, \nonumber
\end{equation}
Given the small thickness of fluids, we ignore inertial effects and gravity. Hence, the equations governing the 
motion of the fluids are Stokes equations for continuity and momentum balance, $ \nabla \cdot {\bf{u}} = 0,\ \nabla \cdot \mathbf{T} = 0,$ respectively,
where $\mathbf{T} = -p \mathbf{I} + 2\eta {\mathbf D}  + \mathbf{\Pi}_M + \Pi_W\, \mathbf{I}$ 
is the total stress tensor and the disjoining pressure is specified by 
\begin{equation}
\label{eq:vdw}
\Pi_W \left( h \right)= \bar \kappa f\left( h \right), \,\,\,\,\mbox{where} \,\,\,\, f\left( h \right) =  \left( {h_*}/{h} \right)^n - \left( {h_*}/{h} \right)^m , \nonumber
\end{equation}
where $\bar \kappa = { \sigma \tan^2 \theta }/{\left(2 M {h}_* \right)}$ and $M = {\left(n - m\right)}/{\left[ \left( m - 1 \right)\left( n - 1 \right) \right]}.$
In this form, $\Pi_W$ includes the disjoining/conjoining intermolecular forces due to van der Waals interactions. 
The prefactor $\bar \kappa $, that can 
be related to the Hamaker constant,  measures the strength of van der Waals forces. Here,  $h_*$ is the short length scale introduced by the van der Waals potential.  We use $n=3$ and $m=2$.     
The contact angle, $\theta$, is the angle at which the fluid/fluid interface meets the substrate.
The addition of the van der Waals forces is crucial in the 
present context, since it allows us to study dewetting of thin films under a magnetic field. 

We nondimensionalizing the governing equations and boundary conditions using the following scales (dimensionless variables are denoted by tilde)
$$
x=x_c \tilde{x}, \,\,\,\,\,\, \left( y, h \right) = h_c \left( \tilde{y},\tilde{h} \right), \,\,\,\,\, \delta = h_c/x_c, \,\,\,\,\,\, u = u_c \tilde{u}, \,\,\,\,\,\,\,\, v=\left( \delta u_c \right) \tilde{v},  \nonumber
$$
$$
p = \left( \eta_2 u_c x_c/h_c^2 \right) \tilde{p}, \,\,\,\,\,\,\, t = \left( x_c/u_c \right) \tilde{t},\,\,\,\,\,\,\,\, \psi = \psi_0 \tilde{\psi},
$$
where $\delta \ll 1$. The initial thickness of the ferrofluid film is denoted $h_c$, and  $u_c$ and $x_c$ are 
the characteristic velocity and horizontal length scale, respectively, given by 
$$
 u_c = \frac{\mu_0\psi_0^2}{\eta_2 x_c}, \,\, x_c = \left( \frac{\sigma h_c^3}{\mu_0\psi_0^2} \right) ^{1/2}. \, \nonumber
$$
Dropping the tilde notation for simplicity, the evolution equation for $h(x,t)$ then becomes
\begin{equation}
\label{eq:axisymetric}
h_t + \frac{1}{3} \frac{1}{x^\alpha} \frac{\partial}{\partial x} \left[ \kappa f'\left(h\right)  x^\alpha h^3 h_{x} - \frac{ \mu_r \left( \mu_r - 1 \right)^{-1}}{\left( h - \frac{\beta \mu_r}{ \mu_r - 1 }\right)^3 } h^3 x^\alpha h_x + x^\alpha h^3 \frac{\partial}{\partial x}\left( \frac{1}{x^\alpha} \frac{\partial}{\partial x} \left(x^\alpha h_x  \right) \right) \right] = 0, 
\end{equation}
where $\kappa={h_c \sigma  \tan^2 \theta }/{\left(2 \mu_0\psi_0^2 M {h}_*\right)}$ is a nondimensional parameter representing 
the ratio of the van der Waals to the magnetic force, 
$\mu_r$ is the ratio of the ferrofluid film permeability to the 
vacuum permeability, $\mu_2/\mu_0$, and  $\alpha=0, 1$ for Cartesian and cylindrical coordinates, respectively.
The ratio $\beta \mu_r/(\mu_r - 1)$ is inversely proportional to the magnetic force (note that $\beta$ is the 
nondimensional distance between the plates with constant potential, 
so the gradient of the potential is inversely proportional to $\beta$).

\begin{figure}
\begin{center}
\includegraphics{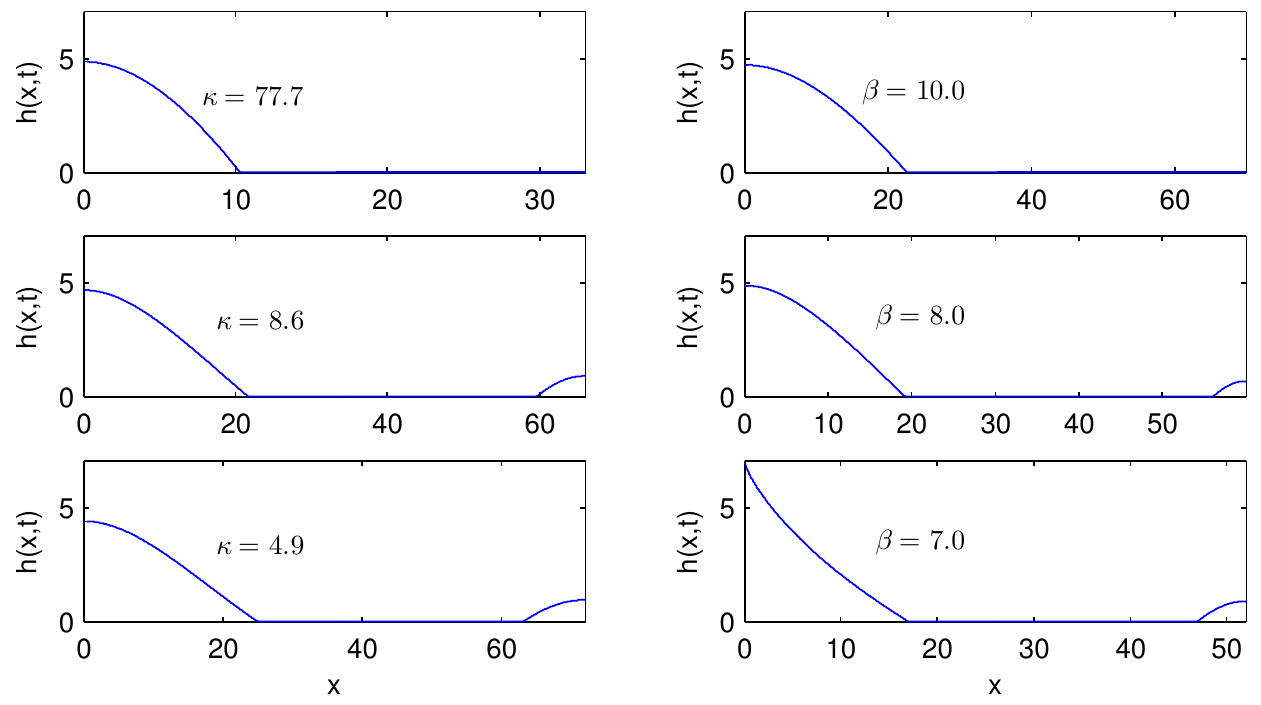}
\end{center}
\hspace{34mm} (a) \hspace{61mm} (b)
  \caption{From \cite{Seric2014}.The effect of varying $\beta$ and $\kappa$ on the film evolution 
                when (a) $\beta = 8$ is fixed, and (b) $\kappa = 13.5$ is fixed. 
                Note that when $\beta \le 7.0$, no steady-state drop profile can be achieved, 
                indicative of an interfacial instability.}
\label{fig:Figure4}
\end{figure}

Here we show the thin film simulations of the steady state profiles 
obtained for a range of parameter values, in particular for nondimensional 
parameters $\beta$ and $\kappa$. We fix $\mu_r = 44.6$ and $h_* = 0.01$.
The initial condition is set to a flat film perturbed around a constant thickness $h_0$, i.e.
$h\left( x, 0 \right) = h_0 + \epsilon \cos{\left( k_m x \right)}\, ,$
with $\epsilon = 0.1$, and $h_0 = 1$.  Here, $k_m$ is the fastest growing mode computed from 
the linear stability analysis in \cite{Seric2014}. 
To put this in perspective,  $\kappa = 13.5$, $\beta = 8$,  and $\mu_r = 44.6$  correspond 
to values of $\sigma$ and $\mu_2$ for an oil-based ferrofluid ($\sigma = 0.034 \mbox{ N/m}$, 
$\mu_2 = \mu_0 (1 + \chi_m)$, where $\chi_m = 3.47 \times 4\pi$), and $\psi_0 = 1.2 \mbox{ A}$, and $\beta$ is chosen to produce a sufficiently strong magnetic field.
Making use of the symmetry of the problem, the computational domain is chosen to be equal to one half of the wavelength of the perturbation, i.e. $ L_x = \pi/k_m$.
No flux boundary conditions are imposed at the left and the right end boundaries as $ h'= h''' = 0 $.
Figure \ref{fig:Figure4}(a) shows the comparison of the steady state profiles for varying parameter $\kappa$ when keeping $\beta$ fixed. We note that as $\kappa$ decreases, i.e. the magnetic force increases compared to the van der Waals force, the satellite
droplets start to appear. Similar behavior is observed in figure~\ref{fig:Figure4}(b) for decreasing values of $\beta$; we observe the formation of satellite 
droplets for sufficiently small $\beta$, i.e. magnetic field dominates over the van der Waals interaction. 
It should also be noted that the satellite droplets are not present in the simulations where the effect of the magnetic field is ignored (i.e.~when $\mu_r = 1 $). 
We also note that for $\beta = 7.0$ shown in figure~\ref{fig:Figure4}(b), a static drop cannot be obtained, unlike when $\beta > 7.0$: here the height of the 
fluid approaches the top boundary and the assumptions of the model are not satisfied for the times later than the one at which this profile is shown.
The results of simulations suggest that  
the satellite droplets form when (i) the magnetic force is sufficiently strong, 
and (ii)  the van der Waals force is sufficiently 
weak, relative to the magnetic force.

%% file: conclusion1.tex
Biomedical technology is expanding the use of coated superparamagnetic nanoparticles and their suspensions in novel directions; for instance, the delivery of drugs to targeted cells, magnetic resonance imaging, early diagnosis of cancer, treatment such as hyperthermia,  and cell separation.  The forefront of research lies at the symbiotic development of nanomedicine, nanotechnology, engineering of strong and highly focused magnets, together with numerical algorithms. In this article, three simplified mathematical models have been presented with the aim of producing numerically generated solutions: stochastic differential equations linked to magnetic drug targeting, a volume-of-fluid computational scheme for the motion of a ferrofluid drop through a viscous medium under a magnetic field, and the evolution of a thin film of ferrofluid with numerical investigation of dewetting. An important future direction is the integration of open-source codes for the numerical simulations used with actual biomedical applications. A drawback of commercial software is that the  location of numerical  inaccuracy in any algorithm is difficult to pinpoint in a black box. The drive toward small scales and complex biomedical domains and materials necessitates a computational approach. 